\documentclass[aps,prd,amsmath,floats,floatfix, twocolumn, superscriptaddress,nofootinbib,showpacs,longbibliography]{revtex4-1}
\usepackage[T1]{fontenc}
\usepackage[utf8]{inputenc}
\usepackage{lmodern}
\usepackage{mathtools}
\usepackage{enumitem}
\usepackage{verbatim}
\usepackage{multirow}
\usepackage{physics}
\usepackage[dvipsnames, usenames]{xcolor}
\definecolor{linkcolor}{rgb}{0.6,0.0,0.0}
\usepackage[hypertexnames=false, unicode, colorlinks=true, linkcolor=linkcolor,
citecolor=linkcolor, filecolor=linkcolor,urlcolor=linkcolor,
pdfusetitle]{hyperref}
\usepackage{orcidlink}

\usepackage[all]{hypcap}
\usepackage{graphicx}
\usepackage{xspace}
\usepackage{amssymb}
\usepackage[normalem]{ulem} %for \sout
\usepackage{bm} % boldmath

\usepackage{microtype}

\usepackage[english]{babel}
\usepackage{blindtext}

\newcommand{\lnlike}{\ln \mathcal{L}}
\def\Re{\mathfrak{\ Re}}

\newcounter{comment}
\begin{document}

\title{Accelerated parameter estimation in Bilby with relative binning}
\author{Kruthi Krishna\orcidlink{0000-0002-6967-5140}}
\affiliation{Department of Physics, Indian Institute of Science, Bangalore 560012, India}
\affiliation{Department of Astrophysics/IMAPP, Faculty of Science, Radboud University, Nijmegen, The Netherlands}

\author{Aditya Vijaykumar\orcidlink{0000-0002-4103-0666}}
\affiliation{Canadian Institute for Theoretical Astrophysics, University of Toronto, 60 St George St, Toronto, ON M5S 3H8, Canada}
\affiliation{International Centre for Theoretical Sciences, Tata Institute of Fundamental Research, Bangalore 560089, India}

\author{Apratim Ganguly\orcidlink{0000-0001-7394-0755}}
\affiliation{Inter-University Centre for Astronomy and Astrophysics (IUCAA), Post Bag 4, Ganeshkhind, Pune 411 007, India}
\affiliation{International Centre for Theoretical Sciences, Tata Institute of Fundamental Research, Bangalore 560089, India}

\author{Colm Talbot}
\affiliation{Kavli Institute for Cosmological Physics, The University of Chicago, 5640 South Ellis Avenue, Chicago, Illinois 60637, USA}

\author{Sylvia Biscoveanu\orcidlink{0000-0001-7616-7366}}
\affiliation{Center for Interdisciplinary Exploration and Research in Astrophysics (CIERA), Northwestern University, 1800 Sherman Ave, Evanston, IL 60201, USA}

\author{Richard N. George}
\affiliation{Weinberg Institute, University of Texas at Austin, Austin, TX 78712, USA}

\author{Natalie Williams\orcidlink{0000-0002-5656-8119}}
\affiliation{School of Physics and Astronomy and Institute for Gravitational Wave Astronomy, University of Birmingham, Edgbaston, Birmingham, B15 2TT, United Kingdom}

\author{Aaron Zimmerman}
\affiliation{Weinberg Institute, University of Texas at Austin, Austin, TX 78712, USA}

\begin{abstract}
We describe an implementation of the relative binning technique to speed up parameter estimation of gravitational-wave signals. 
We first give a pedagogical overview of relative binning, discussing also the expressions for the likelihood marginalized over phase and distance.
Then, we describe the details of the code in \texttt{Bilby}, an open-source software package commonly used for parameter estimation of gravitational-wave sources. Our code is able to reproduce the parameters of GW170817 in 14 hours on a single-core CPU, performs well on simulated signals, and passes the percentile-percentile (p-p) tests. We also illustrate that relative binning is an ideal technique to estimate the parameters of signals in next-generation gravitational wave detectors.
\end{abstract}

\maketitle

\section{Introduction}
Once a compact binary coalescence (CBC) is detected in data from gravitational wave (GW) detectors such as LIGO~\cite{LIGOScientific:2014pky}, Virgo~\cite{VIRGO:2014yos}, and KAGRA~\cite{Somiya:2011np, Aso:2013eba, KAGRA:2020tym}, it is of interest to understand the properties of the CBC that produced the GWs. These source properties include the masses and spins of the objects in the binary, their distance, orientation and sky location with respect to the detector, etc. 
Any further downstream analysis makes use of the properties inferred. These downstream analyses could include, for example, inference of the equation of state of neutron stars~\cite{LIGOScientific:2017ync, LIGOScientific:2018hze, LIGOScientific:2018cki}, population properties of binary black holes~\cite{LIGOScientific:2020kqk, KAGRA:2021duu}, tests of general relativity~\cite{LIGOScientific:2020tif, LIGOScientific:2021sio}, and parameters of an underlying cosmological model~\cite{LIGOScientific:2021aug}.  Therefore, accurate and robust methods of estimating parameters are of utmost importance to GW astronomy. These methods should ideally be quick and computationally efficient---in fact, in certain situations e.g. to permit rapid electromagnetic follow-up of binary neutron star (BNS) mergers or neutron star--black hole (NSBH) mergers, it is desirable to infer the source properties of the neutron stars with a latency of a few minutes.

Bayesian inference is the most favoured methodology for parameter estimation of GW signals. It links a given GW model to the data by constructing posterior distributions for the model parameters. The associated computational cost increases with the dimensionality of the model parameter space and the signal duration. Hence, estimating the source parameters of signals, such as those from BNS events, that last for a long time in ground-based GW detectors is very expensive. Typically, this process takes a few weeks if performed using the most naive Gaussian likelihood prescription. This also does not scale well as we go towards A+~\cite{Miller:2014kma, KAGRA:2013rdx} or the next-generation (XG) GW detectors such as Cosmic Explorer~\cite{LIGOScientific:2016wof} and Einstein Telescope~\cite{Punturo2011}, as they will have longer in-band signals due to increased sensitivity at lower frequencies and higher detection rate given the overall increase in strain sensitivity.

A number of techniques have been developed to reduce the time it takes to estimate parameters. Most of these techniques either concentrate on accelerating the calculation of likelihood or accelerating the overall sampling process. For instance, the reduced order quadrature (ROQ) method~\cite{Canizares:2014fya, Smith:2016qas, Morisaki:2020oqk, Morisaki:2023kuq} speeds up likelihood calculation time by decomposing the waveform parameter space into a set of basis vectors. Since the overlap between the basis vectors can be pre-computed, all that has to be done during sampling is the calculation of the basis coefficients. Other methods of speeding up likelihood computations include multibanded waveform generation~\cite{Vinciguerra:2017ngf}, multibanded likelihood calculation~\cite{Morisaki:2021ngj}, and the meshfree likelihood approximation~\cite{Pathak:2022ktt, Pathak:2023ixb}.
To speed up the sampling process, techniques that rely on parallelized calculations~\cite{Pankow:2015cra, Lange:2018pyp, Smith:2019ucc, Talbot:2019okv}, non-Markovian methods~\cite{Tiwari:2023mzf}, and machine learning methods~\cite{Williams:2021qyt, Green:2020hst} also exist. More recently, there have been proposals to implement waveforms in a ``auto-differentiable'' format to be compatible with \texttt{jax} and harness its power to speed up parameter estimation~\cite{Edwards:2023sak, Wong:2023lgb}.

Relative binning\footnote{Relative binning is often alternatively referred to as the heterodyned likelihood method at various points in the literature. In the rest of the work, we will continue to use relative binning to refer to this method in order to be consistent with the terminology used in the \texttt{Bilby} implementation.}~\cite{Cornish2010, RelBin1, cornish2021} is another way of speeding up the likelihood evaluations. It relies on the approximation that the ratio of neighbouring waveforms in the parameter space of GW signals is smooth, and hence can be approximated by a piecewise linear function. This approximation allows the pre-computation of some terms in the likelihood and reduces the number of frequency points at which the waveform needs to be evaluated during sampling, thus enabling parameter estimation of CBCs (especially neutron stars) in a very short amount of time. In this work, we describe the implementation of the relative binning technique in \texttt{Bilby}~\cite{Ashton:2018jfp, Romero-Shaw:2020owr}, a open-source Bayesian inference package. \texttt{Bilby} provides a user-friendly interface for analyzing GW signals and performing accurate and reliable parameter estimation on both real and simulated data alike~\cite{Ashton:2018jfp, Romero-Shaw:2020owr}. We note that there exist other implementations of relative binning~\cite{Finstad2020, Roulet:2022kot, Narola:2023men, Wong:2023lgb} with varied functionalities, including the generalization to higher harmonics of the GW radiation and precession~\cite{Leslie:2021ssu}, and optimal coordinate systems for sampling~\cite{Roulet:2022kot, Islam:2022afg}; in what follows, we will restrict our discussion to the implementation of relative binning within \texttt{Bilby}, although we shall briefly comment on the validity of our assumptions in Section~\ref{sec:results}.
We also note that an implementation of relative binning compatible with \texttt{Bilby} (but not direclty a part of the \texttt{Bilby} releases) and including extensions to higher modes and precession for a python implementation of the \texttt{IMRPhenomXPHM} waveform is described in Ref.~\cite{Narola:2023men}.

This paper is structured as follows. In Section~\ref{sec:technical overview}, we provide a pedagogical, technical overview of the relative binning technique. In Section~\ref{sec:bilby implementation}, we provide details of the implementation of relative binning in \texttt{Bilby}. In Section~\ref{sec:results}, we describe results from consistency tests of our code with both injections and real events alike. Finally, in Section \ref{sec:summary}, we summarize our results and provide a roadmap for future work. The code described is already available in released versions of \texttt{Bilby}, and has been used to aid parameter estimation in a number of works (e.g. Refs.~\cite{Pradhan:2022rxs,Vijaykumar:2023tjg,Wolfe:2023yuu,Sarin:2023khf,Prakash:2023afe}).

\section{Technical Overview}\label{sec:technical overview}
The key idea behind relative binning is that waveforms with non-negligible posterior
probabilities are alike in the frequency domain and differ only by small perturbations of parameters. A stochastic sampler spends most of its time in a small, high-likelihood neighbourhood around the best-fit waveform. Here the waveforms differ from each other only by small perturbations of parameters (denoted by vector $\bm{\theta})$, and therefore, the ratio between them will be a smooth function \citep{Cornish2010, RelBin1, cornish2021}. Intuitively, by choosing an appropriate set of breakpoints, any smooth function can be approximated by a piece-wise linear function. Let us try to understand how relative binning uses the knowledge of these breakpoints (bin edges) and piece-wise linear coefficients to downsample the required frequency points. The frequency interval between two consecutive breakpoints constitutes a bin $b_i$ (more on the binning criterion in Section \ref{sec:binning}). For a given fiducial waveform $\mu_0(f) = \mu(f, \bm{\theta_{0}})$, and a close-by waveform, $\mu(f, \bm{\theta})$, in the high-likelihood neighborhood, the waveform ratio $r(f) = \mu(f)/\mu_0(f)$ can be approximated by
\begin{equation}\label{eq: r linear}
r(f)\approx
\begin{aligned}
\begin{dcases}
r_{0}({b_1})+r_{1}({b_1})\left(f-f_{\mathrm{m}}({b_1})\right), & f \in b_{1} \\
r_{0}({b_2})+r_{1}({b_2})\left(f-f_{\mathrm{m}}({b_2})\right), & f \in b_{2} \\
\qquad \qquad \vdots
\end{dcases}
\end{aligned}
\end{equation}
where bin $f_m(b_i)$ is the mid-point of the bin $b_i$. Note that, for each waveform, the coefficients $r_{0}({b_i})$ and $r_{1}({b_i})$ are constants in each bin and are independent of frequency.
Moreover, due to the linearity assumption, they can be efficiently computed from the values of $r(f)$ at the bin edges. Hence, we have to evaluate $\mu(f, \bm{\theta})$ only at the bin edges rather than the full frequency array. 

\subsection{Summary Data}\label{sec: sd}

In the relative binning technique, all the information related to the data and the fiducial waveform is condensed into \textit{summary data} and used as weights for the linear-fit coefficients $r_{0}({b_i})$ and $r_{1}({b_i})$ while computing the likelihood. Note that the summary data is independent of the sampled waveform, $\mu(f, \bm{\theta})$. Therefore, it can be evaluated beforehand; for a given event whose parameters need to be estimated, it only needs to be computed once \textit{before} the sampling commences. In this subsection, we will derive the formulae for summary data, following Ref.~\cite{RelBin1}. 

Given the data $d(f)$, the signal duration $T$, the one-sided noise power spectrum density (PSD) of a detector $S_n(f)$, the complex-overlap \citep{Thrane2019, RelBin1} between the data and the waveform template is given by 
\begin{align} 
     Z[d(f), \mu(f , \theta)]
                    &= \sum_{i} \dfrac{4 \ d(f_i) \mu^*(f_i , \theta) } {S_n(f_i) \ T}  \label{eq: Z def} \\
                    &= \sum_{b_j} \biggl( 4 \sum_{f_i \in b_j} \dfrac{d(f_i) \mu^*(f_i , \theta) } {S_n(f_i)  \ T}\biggr) \qq{.} \label{eq: Z def bin}
\end{align}
Here, $x^*$ denotes the conjugate of $x$. In the last step above, we summed over the frequency points in each bin first and then took the sum over all the bins. Now let us substitute $\mu^*(f, \theta) = r^*(f) \mu_{0}^*(f)$ in Eq.\eqref{eq: Z def bin} and simplify the resulting expression as shown below.

% \begin{widetext}
\begin{flalign}
 Z[d, \mu] &=  \sum_{b_j} \biggl( 4 \sum_{f_i \in b_j} \dfrac{d(f_i) r^*(f_i) \, \mu_{0}^*(f_i) } {S_n(f_i) \ T}\biggr) \nonumber \\
                  &\approx \sum_{b_j} \Biggl(  r_{0}^{*}(\mu, {b_j}) \biggl(4 \sum_{f_i \in b_j} \dfrac{d(f_i) \ \mu_{0}^*(f_i) } {S_n(f_i)\ T}\biggr) \nonumber \\
                & + r_{1}^{*} (\mu, {b_j}) \biggl(4 \sum_{f_i \in b_j} \dfrac{d(f_i) \ \mu_{0}^*(f_i ) } {S_n(f_i)\ T} (f_i-f_{\mathrm{m}}({b_j})) \biggr) \Biggr) \nonumber\\
                &\approx \sum_{{b_j}}\left(A_{0}({b_j}) r_{0}^{*}(\mu, {b_j})+A_{1}({b_j}) r_{1}^{*}(\mu, {b_j})\right)     \label{eq: Zdh derivation}\\
\mathrm{where} & \quad A_{0}({b_j}) =4 \sum_{f \in {b_j}} \frac{d(f) \mu_{0}^{*}(f)}{S_{n}(f) \ T} \nonumber\\
& \quad A_{1}({b_j}) =4 \sum_{f \in {b_j}} \frac{d(f) \mu_{0}^{*}(f)}{S_{n}(f) \ T}\left(f-f_{\mathrm{m}}({b_j})\right) \nonumber
\end{flalign}
% \end{widetext}
Similarly, we can derive equivalent expressions for $Z[\mu(f), \mu(f)]$. Thereby, we get the following formulae for summary data -- $A_{0}({b_j})$, $A_{1}({b_j})$, $B_{0}({b_j})$, $B_{1}({b_j})$ - and complex-overlaps: 
\begin{equation}\label{eq: summary-data}
\begin{aligned} 
A_{0}({b_j}) &=4 \sum_{f \in {b_j}} \dfrac{d(f) \mu_{0}^{*}(f)}{S_{n}(f) \ T} \\
A_{1}({b_j}) &=4 \sum_{f \in {b_j}} \dfrac{d(f) \mu_{0}^{*}(f)}{S_{n}(f) \ T}\left(f-f_{\mathrm{m}}({b_j})\right) \quad\\
B_{0}({b_j}) &=4 \sum_{f \in {b_j}} \dfrac{\left|\mu_{0}(f)\right|^{2}}{S_{n}(f) \ T} \\
B_{1}({b_j}) &=4 \sum_{f \in {b_j}} \dfrac{\left|\mu_{0}(f)\right|^{2}}{S_{n}(f) \ T}\left(f-f_{\mathrm{m}}({b_j})\right)
\end{aligned}
\end{equation}
\begin{align}
Z[d(f), \mu(f)] &\approx \sum_{{b_j}}\left(A_{0}({b_j}) r_{0}^{*}(\mu, {b_j})+A_{1}({b_j}) r_{1}^{*}(\mu, {b_j})\right) \label{eq: Zdh} \\
Z[\mu(f), \mu(f)] &\approx \sum_{{b_j}} \biggl(B_{0}({b_j})\left|r_{0}(\mu, {b_j})\right|^{2} \nonumber\\ & \qquad +2 B_{1}({b_j}) \Re \left[r_{0}(\mu, {b_j}) r_{1}^{*}(\mu, {b_j})\right]\biggr)  \label{eq: Zhh}
\end{align}

In Eq.~\eqref{eq: Zdh} and \ref{eq: Zhh}, the complex-overlaps are expressed as functions of terms that are only dependent on the bin, not the frequency itself.
Thus, the number of frequency points required to evaluate $Z[d(f), \mu(f)]$ and $Z[\mu(f), \mu(f)]$ is reduced.

For the sake of completeness, we write down the formula used for computing the likelihood below.
\begin{flalign}\label{eq: ln_def2}
\lnlike(\bm{\theta}) =& \log \mathcal{Z_N} + \Re{Z[d(f), \mu(f, \bm{\theta})]} \nonumber\\& - \dfrac{1}{2}Z[\mu(f, \bm{\theta}), \mu(f, \bm{\theta})] 
\end{flalign}
where $\mathcal{Z_N}$ is the noise evidence \citep{Thrane2019}. Tables \ref{tab:fvsbin} and \ref{tab:dependency} summarise all the dependencies of various terms we have encountered so far.

\begin{table}
\setlength\tabcolsep{2mm}
\caption{A comparison between quantities evaluated at each frequency point and those only evaluated at bin edges.}{} \label{tab:fvsbin} 
\begin{tabular}{c|c|c} 
\toprule
\multirow{2}{*}{\textbf{Quantity}}  & \multicolumn{2}{c}{Evaluated at} \\ 
\cline{2-3} & \textbf{each $f$ point} & \textbf{bin edges} \\ 
\hline
Data ($d$) & \checkmark & \\ 
\hline
Fiducial waveform $(\mu_0)$ &\checkmark &  \\ 
\hline
A sampled waveform  $(\mu)$ & &  \checkmark \\ 
\hline 
Waveform ratio $(r = \mu/\mu_0)$ & & \checkmark \\ 
\toprule
\end{tabular} 
\end{table}

\begin{table}
\setlength\tabcolsep{4mm}
\caption{Relationship between summary data, linear-fit coefficients, data, fiducial waveform, and sampled waveform.}{} \label{tab:dependency} 
\begin{tabular}{c|c|c|c} 
\toprule
\multirow{2}{*}{\textbf{Quantity}} & \multicolumn{3}{c}{Depends on}\\ 
\cline{2-4} 
& $\mathbf{d}$ & $\mathbf{\mu_0}$ & $\mathbf{\mu}$ \\ 
 \hline
Summary data & \checkmark & \checkmark &   \\ 
\hline
linear-fit coefficients $r_0$ and $r_1$ &  & \checkmark & \checkmark \\ 
\toprule
\end{tabular} 
\end{table}

\subsection{Phase and Distance Marginalized Likelihoods}
\label{subsec:phase-dist}

In GW parameter estimation, to speed up the calculations and improve convergence, we often explicitly marginalize over extrinsic parameters such as coalescence time,  phase ($\phi$)~\cite{Veitch2015} and luminosity distance ($D_L$) \citep{Thrane2019}. In Appendix \ref{sec:time-marg}, we argue that marginalizing over coalescence time is not feasible with relative binning. However, the prescription given in \citet{Thrane2019} for likelihoods marginalized over $\phi$ and $D_L$ remains the same with relative binning. We rewrite the marginalized likelihoods in complex-overlap notation below:
\begin{align}
 \log \mathcal{L}_\text{marg}^\phi (\bm{\tilde{\theta}}) =& \log{\cal Z}_N - \dfrac{1}{2}Z[\mu(\bm{\tilde{\theta}}), \mu(\bm{\tilde{\theta}})] \nonumber \\& + \log I_0\Big(\Big|Z[d, \mu(\bm{\tilde{\theta}})]\Big|\Big) \label{eq: phase-marge-relbin}\\
 \log \mathcal{L}_\text{marg}^D (\bm{\tilde{\theta}}) =& \log \mathcal{Z}_N + \log {\cal L}_D(\kappa^{2}(\bm{\tilde{\theta}}),\rho_\text{opt}(\bm{\tilde{\theta}})) \label{eq: dist-marg-relbin}
 \end{align}
 where in each equation, $\bm{\tilde{\theta}}$ is the set of binary parameters excluding the parameter we are marginalizing over, and  $I_0$ is the modified Bessel function of first kind. The complex overlaps, $\kappa^{2} = \Re(Z[d, \mu])$ and $\rho_\text{opt} = Z[\mu, \mu]$, are employed in pre-computing a lookup table for $\mathrm{log} {\cal L}_D(\kappa^{2},\rho_\text{opt})$ using the integral below.
\begin{align}
{\cal L}_D(\kappa^{2},\rho_\text{opt}) &\equiv \int d D_{L}\, \exp \biggl(\Re(Z[d, \mu(\bm{\tilde{\theta}}, D_{L})]) \nonumber \\ & - \dfrac{1}{2} Z[\mu(\bm{\tilde{\theta}}, D_{L}), \mu(\bm{\tilde{\theta}}, D_{L})]\biggr) \pi(D_{L}) .
\end{align}
where $\pi(D_{L})$ is the prior distribution of $D_L$.

\subsection{Binning Criterion}\label{sec:binning}
Relative binning likelihood is an approximation to the exact likelihood. The likelihood error largely depends on the accuracy of the piece-wise linear approximation of the waveform ratio, $r(f)$. Thus, the objective is to choose a minimal set of breakpoints that fully capture the variations in $r(f)$ and yield a likelihood error within a given tolerance level. We follow \citet{RelBin1} in this work. However, an alternate way to do this is outlined in Ref~\cite{cornish2021}.

% \apratim{it will be good to mention the alternate method in a couple of sentences, if possible}. I don't understand Cornish's method, so leaving it out - Kruthi

For small perturbations in parameters, $\Delta\bm{\theta}$, by expressing a waveform $\mu(f)$, in terms of its phase $\Psi(f)$ and magnitude $|\mu(f)|$, the waveform ratio becomes
\begin{equation}
\begin{aligned} 
&r(f) = \left|\dfrac{\mu(f)}{\mu_0(f)}\right|e^{i\Delta_{\theta}\Psi(f)} \\[4pt]
&\mathrm{where}\enspace\Delta_{\theta}\Psi(f) = \Psi(f,\,\bm{\theta_{0}} + \Delta\bm{\theta}) - \Psi(f,\,\bm{\theta_{0}})
\end{aligned}
\end{equation}
In the equation above, the magnitude term is slow-varying compared to the phase-difference term, $\Delta_{\theta}\Psi(f)$. Thus, the oscillations of $r(f)$ are mainly characterized by $\Delta_{\theta}\Psi(f)$. This allows us to choose the breakpoints based on the criterion that the change in $\Delta_{\theta}\Psi(f)$ over a frequency bin $b_n$ is some small number $\epsilon$, in radians,
\begin{equation}\label{eq: binning-criterion-theta-dependent}
\bigl|\Delta_{\theta} \Psi(f_{max}(b_n)) - \Delta_{\theta} \Psi(f_{min}(b_n))\bigr| = \epsilon.
\end{equation}
However, note that $\Delta_{\theta}\Psi(f)$ is dependent on $\bm{\theta} = \bm{\theta_{0}} + \Delta\bm{\theta}$, which means our breakpoints will be different for each sampled waveform. Let us examine how we can get rid of this $\bm{\theta}$-dependence in our binning criterion.

In the post-Newtonian theory, $\Psi(f )$ can be formulated as a sum of different powers of $f$,
\begin{equation}\label{eq: phase-sum}
\Psi(f, \bm{\theta}) = \sum_{k} \alpha_k (\bm{\theta}) f^{\gamma_k}, 
\end{equation}
where $\alpha_k$'s are the power-law coefficients and $\gamma_k$'s are the power-law indices \cite{Blanchet:2013haa}. The phase-difference $\Delta_{\theta}\Psi(f)$ does not vary by more than a few cycles for small perturbations $\Delta\bm{\theta}$. In this case, we can impose a limit on the maximum perturbation of each term in Eq.~\eqref{eq: phase-sum}, $|\Delta \alpha_k f^{\gamma_k}|_{max} \approx 2 \pi \chi$, where $\chi$ is a tunable factor. For a frequency range $[f_{\mathrm{low}}, f_{\mathrm{high}}]$, this gives us
\begin{equation}\label{eq: key}
\begin{aligned}
&|\Delta \alpha_k^{\max}| f^{\gamma_k}_{k_*} \approx 2 \pi \chi \\
&\text{where}\enspace f_{k_*}  = \begin{dcases}
f_{\mathrm{low}} &\text{if  }  \gamma_k < 0\\
f_{\mathrm{high}} &\text{if  }  \gamma_k > 0
\end{dcases}
\end{aligned}
\end{equation}
The above equation gives us the maximum perturbation allowed for each $\alpha_k$ in our sampling scenario. Therefore, the phase-difference corresponding to the maximum perturbation of $\alpha_k$'s will be 
\begin{equation}\label{eq: phi-theta-independent}
\begin{aligned}
\Delta_{\theta} \Psi'(f) &=  \sum_{k} \Delta\alpha_k^{\max}\ f^{\gamma_k}\  \\
			    &= 2 \pi \chi \sum_{k}  \left(\dfrac{f}{f_{k_*}}\right)^{\gamma_k} \mathrm{sgn}(\gamma_k).
\end{aligned} 
\end{equation}
In the last step above, the sign function, $\mathrm{sgn}(\gamma_k)$, is added to each term in the summation to allow for a scenario where signs of $\alpha_k$'s conspire to generate the maximum phase difference. We shall use this $\bm{\theta}$-independent estimate of phase difference in Eq.~\eqref{eq: binning-criterion-theta-dependent} to obtain our binning criterion:
\begin{equation}\label{eq: binning-criterion-theta-independent}
f_{max}(b_n)= \Delta_{\theta} \Psi'^{-1}(n\epsilon + \Delta_{\theta} \Psi'(f_{\mathrm{low}}))
% \bigl|\Delta_{\theta} \Psi(f_{max}(b)) - \Delta_{\theta} \Psi(f_{min}(b))\bigr| = \epsilon.
\end{equation}
where $n = 1, 2, ...,$ is the bin number and $f_{\mathrm{low}}$ is the minimum frequency cut-off of our signal. 

\subsection{Pre-sampling}
For Bayesian parameter estimation with relative binning, prior to the familiar, stochastic sampling, the following pre-sampling steps are necessary:
\begin{enumerate}
\item \textit{Setting up bins}: An efficient way to implement Eq.~\eqref{eq: binning-criterion-theta-independent} is to divide the range $[\Delta_{\theta} \Psi'(f_{\mathrm{low}}), \Delta_{\theta}\Psi'(f_{\mathrm{high}})]$ in a grid of spacing $\epsilon$ and invert the grid points to obtain $f_{max}(b_n)$.

\item \textit{Fixing the fiducial waveform}: Reasonable choices for fiducial parameters ($\bm{\theta_0}$) would be the best-fit parameters outputted by matched filter searches, the maximum likelihood parameters or injection parameters in case of an injected signal \citep{RelBin1, RelBin2}. 

\item \textit{Computing summary data}: With the knowledge of the data, fiducial waveform, bins, and the noise PSD, we can obtain the summary data for each detector using Eq.~\eqref{eq: summary-data}.
\end{enumerate}

\subsection{Sampling}\label{sec: sampling}
Posterior distributions for a chosen set of parameters are obtained by stochastic sampling different waveforms in the parameter space. To evaluate the corresponding likelihood value for each sampled waveform, we execute the following steps:
\begin{enumerate}
\item \textit{Evaluate $\mu(f)$}: Here the waveform is generated only for the bin edges rather than the full frequency grid. This immensely reduces the overall waveform generation runtime.
\item \textit{Calculate $r(f)$}: At each bin edge, we calculate the ratio between the sampled waveform $\mu(f)$ and the fiducial waveform $\mu_0(f)$.
\item \textit{Obtain $r_0$ and $r_1$}: Using the waveform ratio values at the bin edges, the linear-fit coefficients $r_0$ and $r_1$ can be obtained for each bin with simple algebra.
\item \textit{Compute the complex-overlaps and the likelihood}: In the final step, we calculate the complex-overlaps and the likelihood using Eqs.~\eqref{eq: Zdh}, \eqref{eq: Zhh}, and \eqref{eq: ln_def2}. 

\end{enumerate}

\section{Implementation in Bilby}\label{sec:bilby implementation}
Our implementation of the relative binning technique has been publicly available \texttt{Bilby} \cite{Ashton:2018jfp} since version 1.4.0. We added a new likelihood subclass called \texttt{RelativeBinningGravitationalWaveTransient} which, once setup, internally calls different functions that carry out various steps described in the previous sections. We outline a few important functions below:
\begin{itemize}
\item \texttt{setup\textunderscore bins} computes the $\Delta_{\theta} \Psi'$ for the full frequency array that \texttt{waveform\textunderscore generator} generates using five power-law indices $\gamma_k = \{-5/3, -2/3, 1, 5/3, 7/3\}$ \citep{RelBin1} corresponding to the five leading-order post-newtonian terms (see e.g. Ref.~\cite{Buonanno:2009zt}), then divides it into a $\epsilon$-spaced grid and inverts those $\Delta_{\theta} \Psi'$-- grid points to obtain our frequencies bin edges. Thereby, the number of $\Delta_{\theta} \Psi'$-- grid points determines the total number of bins used in the analysis.
\item \texttt{set\textunderscore fiducial\textunderscore waveforms} evaluates the fiducial waveform values for the full frequency array using the input fiducial parameters. However, if no user-input fiducial parameters are available, a set of parameters is randomly drawn from the prior. We also allow the user to input an initial guess for the fiducial parameters and refine it further using a pre-computed optimization technique by invoking \texttt{update\textunderscore fiducial\textunderscore parameters=True}. This is done using by running a few iterations of a \texttt{scipy.optimize}~\cite{2020NatMe..17..261V} routine called through a helper function \texttt{find\textunderscore maximum\textunderscore likelihood\textunderscore parameters}.
\item \texttt{compute\textunderscore summary\textunderscore data} computes and stores the summary data for each interferometer used.
\item \texttt{compute\textunderscore waveform\textunderscore ratio\textunderscore per\textunderscore interferometer} evaluates the waveforms at bin edges and calculates the waveform ratio using the already stored fiducial waveform for each interferometer.
\item \texttt{calculate\textunderscore snrs} computes the complex overlaps (equation \ref{eq: Zdh} and \ref{eq: Zhh}) for each interferometer, using the summary data and waveform ratios.
\end{itemize}
Furthermore, we have introduced new source models \texttt{lal\textunderscore binary\textunderscore neutron\textunderscore star\textunderscore relative\textunderscore binning} and \texttt{lal\textunderscore binary\textunderscore black\textunderscore hole\textunderscore relative\textunderscore binning} that allow waveform generation at bin edges rather than the full frequency array. However, they can be set to utilize the full frequency array when generating an injected signal\footnote{An example for performing parameter estimation with relative binning on an injected signal can be found in the \texttt{Bilby} git repository at \href{https://git.ligo.org/lscsoft/bilby/-/blob/7afbc7479c81b30c5a163f3169734c965e3bd623/examples/gw_examples/injection_examples/relative_binning.py}{this URL}.} by adding an additional parameter \texttt{fiducial=1} to the injection parameter dictionary. 

We note that, by implementing this likelihood in \texttt{Bilby}, we have extended the implementation of relative binning used in \citet{RelBin1} to incorporate a fully coherent network statistic and include extrinsic parameters such as distance, sky location, inclination, and polarization (also see \cite{Finstad2020, Raaijmakers2021}), and enabled marginalization over the coalescence phase and luminosity distance as described in subsection \ref{subsec:phase-dist}. Furthermore, in appendix \ref{sec:time-marg}, we argue that time marginalization is not practical for relative binning in \texttt{Bilby}.

\section{Results}\label{sec:results}
\begin{figure*}[!htbp]
   \centering
   \includegraphics[width = 16.5cm]{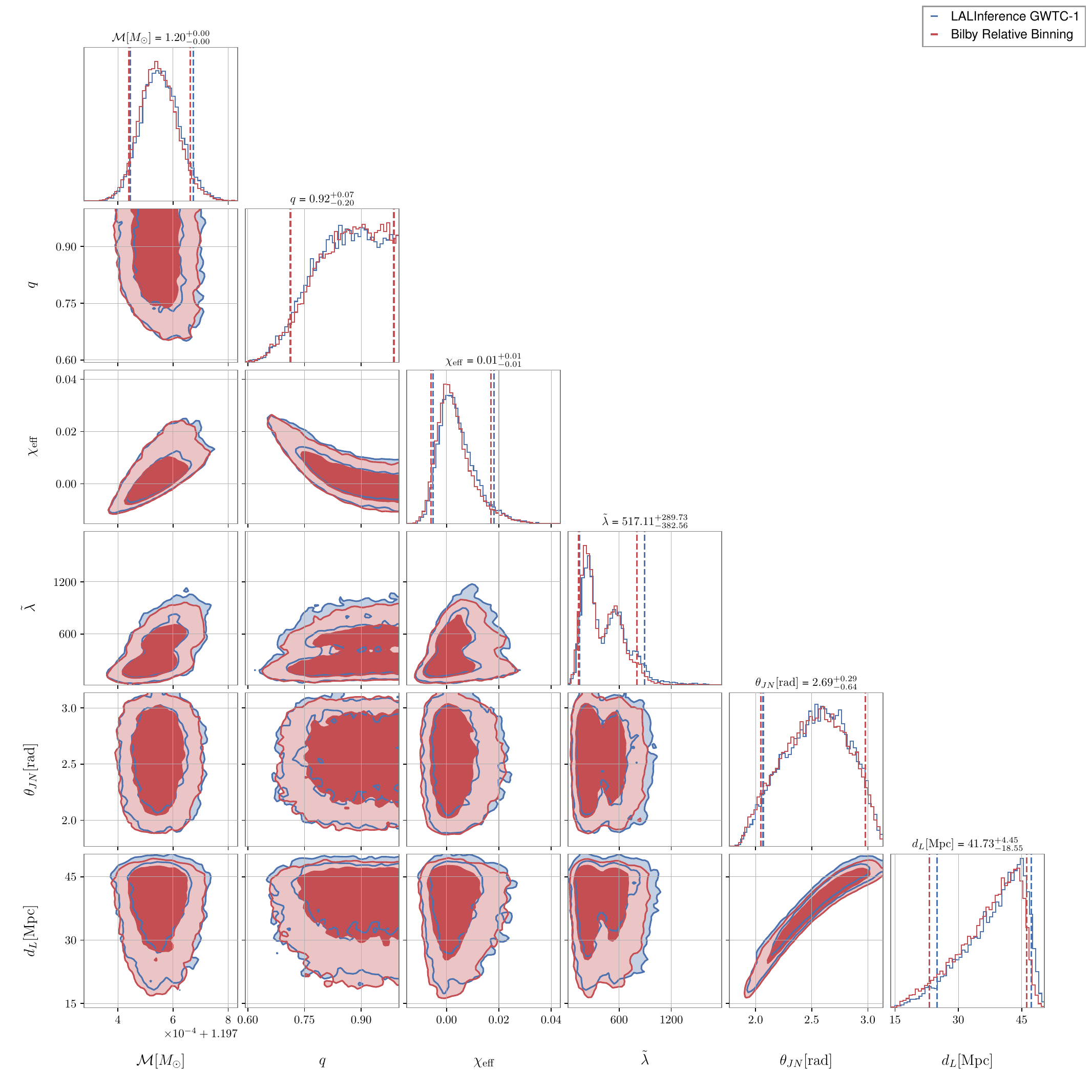} 
      \caption{Comparison of posterior distributions from a relative binning parameter estimation of GW170817 in \texttt{Bilby} (red) as compared to \texttt{LALInference} GWTC-1 samples obtained using the fulllikelihood computation (blue).
      Marginalized 1-dimensional histograms for each parameter are shown along the diagonal with dashed vertical lines and labels for 5\% and 95\% quantiles of the parameters.
      Off-diagonal plots show 2-dimensional joint-posterior distributions with contours corresponding to the 68\% and 95\% credible regions.}
      \label{fig: gw170817}
\end{figure*}

In this section, we first compare the performance of relative binning against the exact method in terms of accuracy and computational costs for both real and simulated signals.
Then, we explore the accuracy of the current implementation of relative binning for non-quadrupolar waveforms and its use for next-generation detectors. Unless otherwise specified, the priors are uniform in $\sin \delta$ where $\delta$ is declination, uniform in $\cos \theta_{JN}$ where $\theta_{JN}$ is the orientation angle, and uniform in right ascension $\alpha$, polarization angle $\psi$, coalescence phase $\phi_c$ in their respective ranges. All analyses are performed with v2.1.2 of \texttt{Bilby} and v1.1.2 of \texttt{Bilby\textunderscore pipe}. We use the \texttt{Dynesty} sampler~\cite{Speagle:2019ivv} as implemented in \texttt{Bilby} with the custom \texttt{acceptance-walk} stepping method (nlive=1000, naccept=60) to sample over relevant parameters.  

\subsection{GW170817 Parameter Estimation}

BNS events like GW170817~\cite{LIGOScientific:2017vwq} span the entire range of detectable frequencies of LIGO and Virgo detectors and require a fine sampling of parameter space which makes their parameter estimation computationally much more expensive than a typical binary black hole (BBH). To demonstrate the effectiveness of our code in achieving both accuracy and computational efficiency for such events, we analyzed 256s duration of GW170817 strain data from LIGO Livingston, LIGO Hanford, and the Virgo detectors sampled at 4096 Hz. Our analysis covered a frequency range of 20Hz to 2048Hz, and the sampling time was approximately 14 hours with a single CPU core. We expect the sampling time to improve when parallelized over multiple cores, readily doable with \texttt{Bilby}. The initial parameter estimates given in \citet{LIGOScientific:2017vwq} were taken as initial guesses for the fiducial parameters but were further updated to the maximum likelihood estimates by setting \texttt{update\textunderscore fiducial\textunderscore parameters=True}. To set up the bins,  we take $\epsilon$ = 0.25 radians (or alternatively $\chi$ = 2) informed by the discussions in \citet{RelBin1}, which results in 123 frequency bins with our code. We use the \texttt{IMRPhenomPv2\_NRTidal} model \cite{Dietrich2019} and the \texttt{Dynesty} sampler~\cite{Speagle:2019ivv} coupled with \texttt{Bilby} for our analysis.

In order to compare our posteriors to the readily available GWTC-1~\cite{LIGOScientific:2018mvr} samples\footnote{\url{https://dcc.ligo.org/LIGO-P1800370/public}} we have fixed the sky location (right ascension and declination) to the known position of SSS17a/AT 2017gfo as determined by electromagnetic observations \cite{LIGOScientific:2017ync, GWTC-1}.
We adopt the low-spin priors, meaning the spin magnitudes of the two neutron stars $a_1$ and $a_2$ are drawn uniformly from the interval [0, 0.05], while the spin orientation angles are drawn isotropically. The prior on mass ratio $q$ is also assumed to be flat between [0.125, 1]. To speed up convergence, we impose a flat prior [1.18, 1.21] $M_\odot$ on the detector frame chirp mass $\mathcal{M}$. We notice that, in general, relative binning performs well when we restrict $\mathcal{M}$ to a small region centred on the fiducial parameter $\mathcal{M}_0$. To be consistent with the prior choices employed in the GWTC-1 result, we reweight the posteriors obtained with a flat prior on chirp mass and mass ratio to a flat prior on the individual component masses. Figure \ref{fig: gw170817} shows a comparison of posteriors between relative binning and the GWTC-1 samples, which were obtained using the full likelihood calculation.
The posteriors obtained from both the methods show excellent agreement with each other. To quantify the ``closeness'' between the two sets of posterior samples, we use the Jensen-Shannon (JS) divergence~\cite{Lin:1991zzm}. Specifically, we calculate the JS divergence for each parameter that we sampled over, and quote the maximum value of JS divergence across parameters as our similarity statistic. Following Ref.~\cite{Narola:2023men}, we choose a threshold of 0.06 for the JS divergence. For GW170817, the maximum JS divergence across parameters is 0.007 (corresponding to $D_L$), well below our threshold.

\begin{table*}[htb!]
\caption{Injection parameters used in the results from Section~\ref{sec:injections}, \ref{sec:hm} and \ref{sec:xg}.}

\begin{tabular}{|l|l|l|l|l|}
    \hline
        \textbf{Parameter} &  \textbf{BBH}  &  \textbf{BBH HM }         & \textbf{BBH XG}        &\textbf{  BNS XG}        \\ \hline
        Total detector-frame mass $M_{\rm total} / M_{\odot}$  & 70 & 60 & 68.92 & 5.51  \\ \hline
        Mass Ratio $q$  & 0.8 & 0.5 & 1 & 1  \\ \hline
        Primary spin magnitude $a_1$  & 0.01 & 0.5 & 0 & 0 \\ \hline
        Secondary spin magnitude $a_2$  & 0.02 & 0.5 & 0 & 0 \\ \hline
        Primary spin tilt $\theta_1$  & 0 & 1 & 0 & 0  \\ \hline
        Secondary spin tilt $\theta_2$  & 0 & 0.5 & 0 & 0  \\ \hline
        Azimuthal angle between the two component spins $\phi_{12}$  & 0 & 1.7 & 0 & 0  \\ \hline
        Angle between total and orbital angular momenta $\phi_{{JL}}$  & 0 & 0.3 & 0 & 0  \\ \hline
        Right ascension $\alpha$  & 1.67 & 2.5 & 1.67 & 1.67 \\ \hline
        Declination $\delta$  & -1.26 & 0 & -1.26 & -1.26  \\ \hline
        Geocentric time $t_{c} / \mathrm{s}$  & 1126259462 & 1126259642 & 1126259462 & 1126259462  \\ \hline
        Luminosity Distance $D_L / \mathrm{Mpc}$  & 701.58 & 833.54 & 7000 & 701.58  \\ \hline
        Polarization angle $\psi$  & 3.93 & 0 & 3.9 & 3.93 \\ \hline
        Orientation angle $\theta_{{JN}}$  & 2.82 & 1.04 & 2.82 & 2.82 \\ \hline
        Reference phase $\phi$  & 0 & 0 & 0 & 0 \\ \hline
    \end{tabular}
    \label{tab:injection-parameters}
\end{table*}

\subsection{Injection Study}
\label{sec:injections}

\begin{figure}[]
   \centering
    \includegraphics[width=\linewidth]{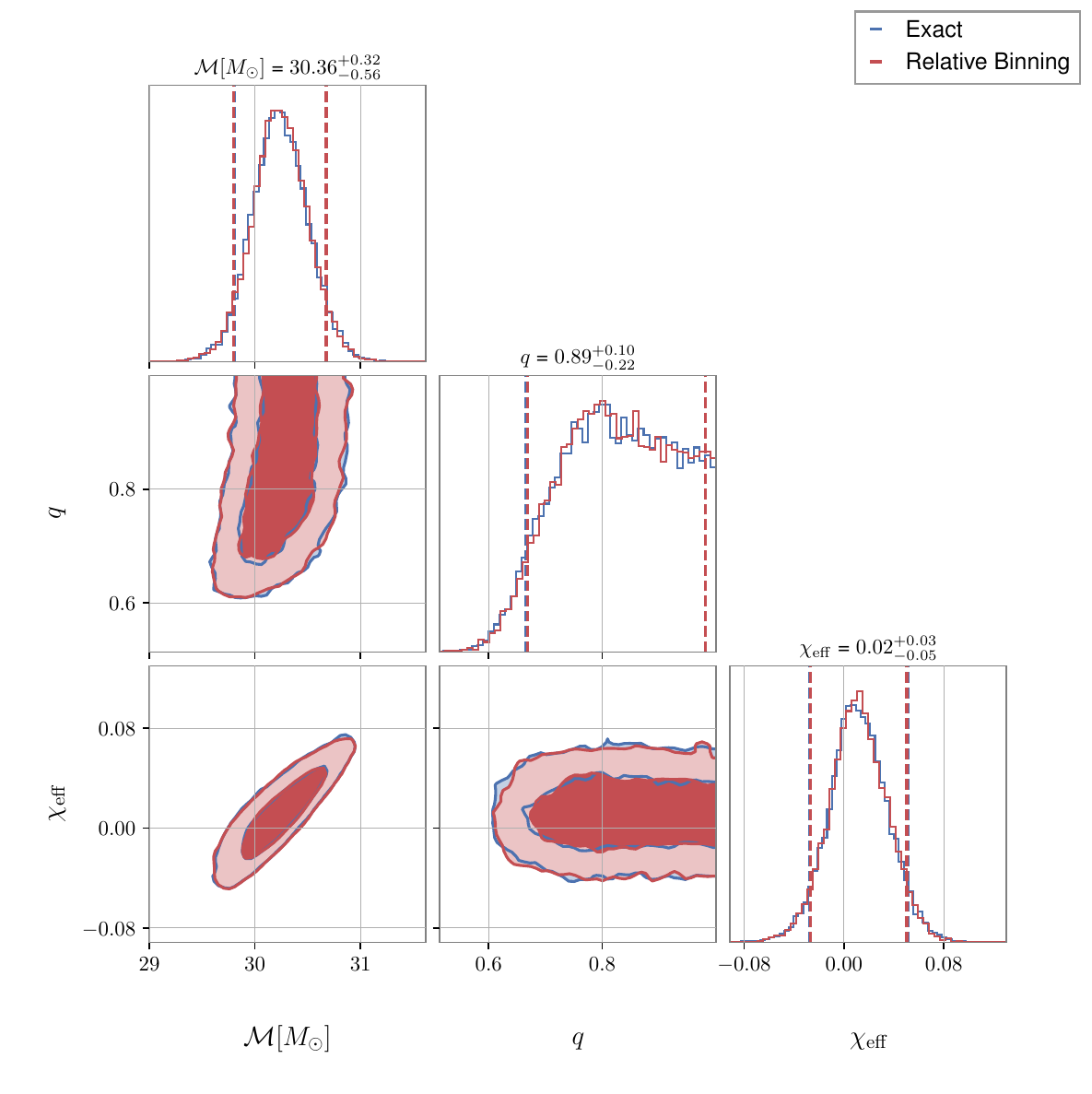} 
    \caption{Comparison of posterior distributions for intrinsic parameters of a BBH injection estimated using relative binning (red) and exact (blue) methods in \texttt{Bilby}.
      Marginalized 1-dimensional histograms for each parameter are shown along the diagonal with dashed vertical lines and labels for 5\% and 95\% quantiles of the parameters.
      Off-diagonal plots show 2-dimensional joint-posterior distributions with contours corresponding to the 68\% and 95\% credible regions. The waveform approximant used is \texttt{IMRPhenomPv2}.}
      \label{fig:bbh_inj_intrinsic}
\end{figure}

\begin{figure*}[!htbp]
   \centering
   \includegraphics[width=15cm]{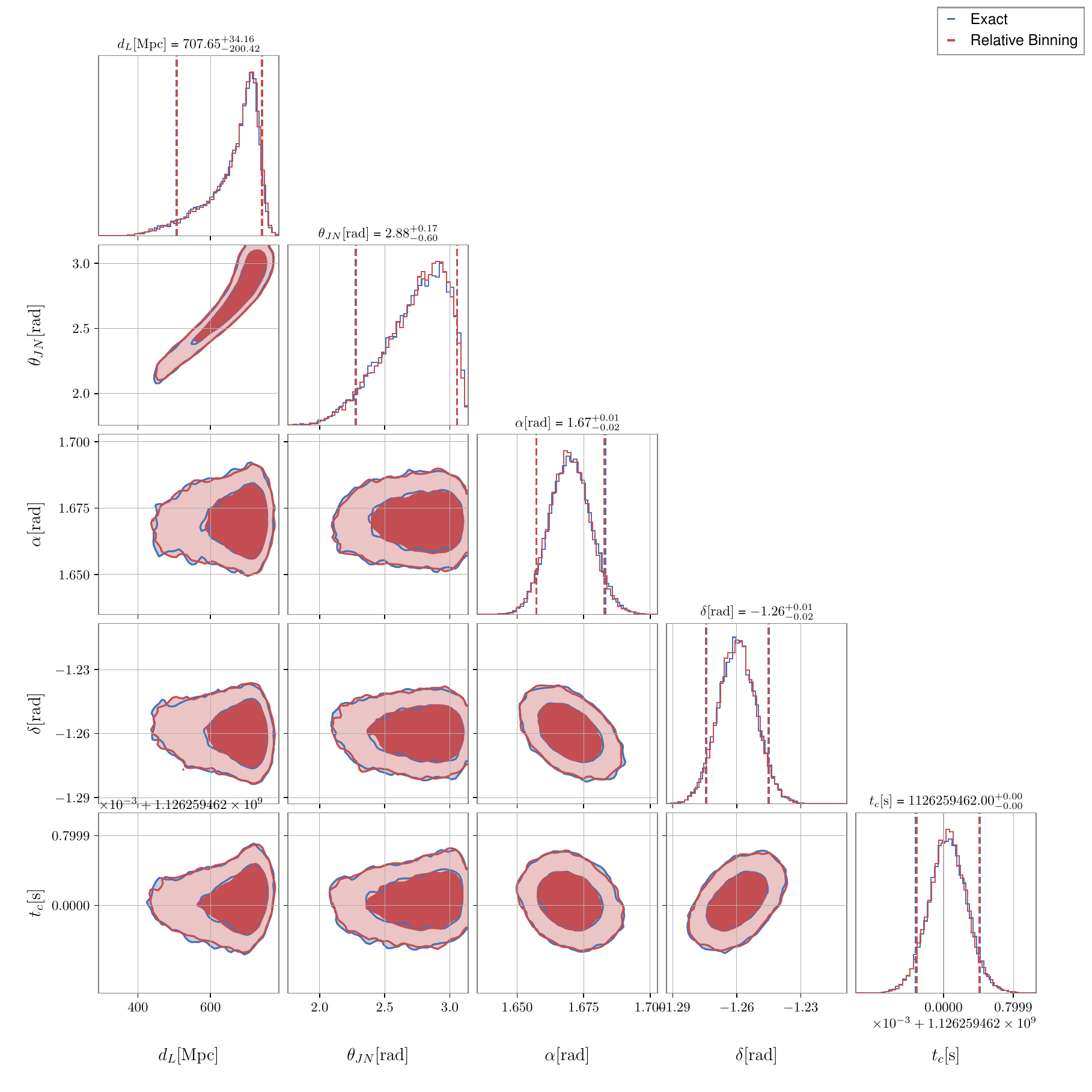} 
 \caption{Comparison of posterior distributions for extrinsic parameters of the BBH injection seen in Figure \ref{fig:bbh_inj_intrinsic}, estimated using relative binning (red) and exact (blue) methods in \texttt{Bilby}.
 Marginalized 1-dimensional histograms for each parameter are shown along the diagonal with dashed vertical lines and labels for 5\% and 95\% quantiles of the parameters.
 Off-diagonal plots show 2-dimensional joint-posterior distributions with contours corresponding to the 68\% and 95\% credible regions.
 The waveform approximant used is \texttt{IMRPhenomPv2}.}
      \label{fig:bbh_inj_extrinsic}
\end{figure*}

In an injection study, a synthetic GW signal is added to the simulated data and the resulting data stream is analyzed. A brief discussion on how to do this in \texttt{Bilby}\footnote{An example code can be found here: \url{https://lscsoft.docs.ligo.org/bilby/basics-of-parameter-estimation.html}} can be found in Ref.~\cite{Ashton:2018jfp}. To assess the performance of our code, we injected non-precessing BBH and BNS signals in typical regions of the parameter space relevant to astrophysical scenarios and performed a full parameter estimation. The injected signals had network signal-to-noise ratio (SNR) $ \geq 12$. We took the injection parameters to be the fiducial parameters and used 123 frequency bins ($\epsilon$ set to 0.25 radians). We repeated the same analysis with the exact likelihood calculation in \texttt{Bilby} for the BBH signals. We use the zero-noise realization in both cases in order to allow for a one-to-one comparison between the two methods. Here, we used the waveform approximant \texttt{IMRPhenomPv2}~\cite{Khan:2018fmp} and sampled all the 15 parameters associated with a BBH merger. Figures \ref{fig:bbh_inj_intrinsic} and \ref{fig:bbh_inj_extrinsic} show a comparison of the posteriors from the two methods for an example BBH signal; they agree well with each other, with the maximum JS divergence being 0.004. The injection parameters used in this example are outlined in Table \ref{tab:injection-parameters}. The injection was recovered using isotropic priors on spins with spin magnitudes between 0 and 0.99,  chirp mass prior uniformly distributed over [20, 60]$M_\odot$, and luminosity distance $D_L$ prior distributed as $\propto D_L^2$ over the range 10 Mpc to 5000 Mpc.

\begin{figure}[!htbp]\label{fig: BNS pp-test}
    \includegraphics[width=\linewidth]{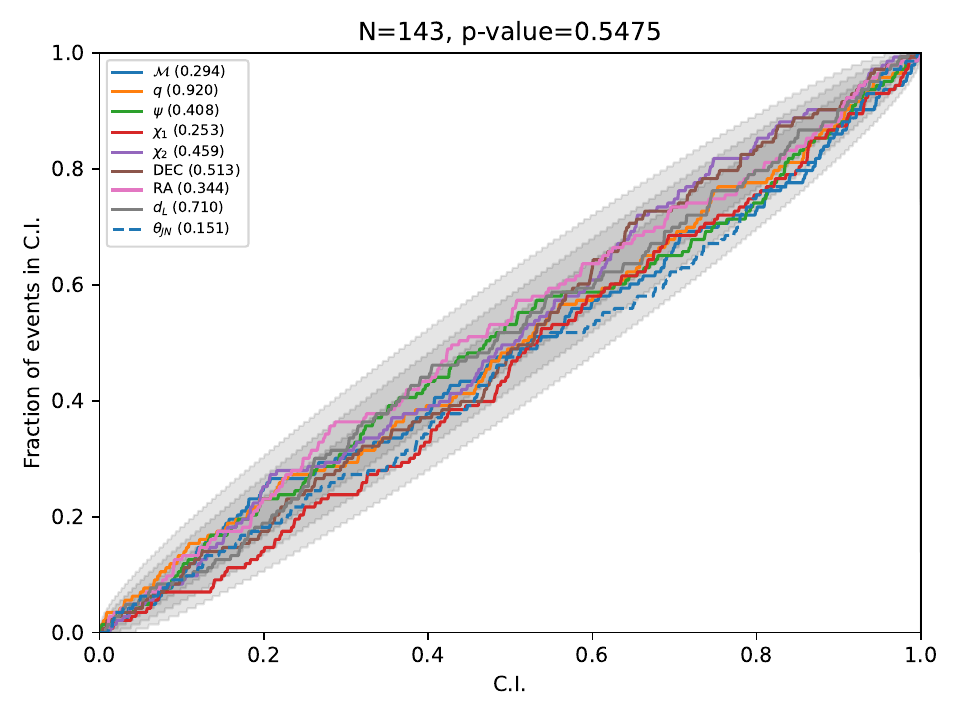}
    \caption{Percentile-percentile (p-p) plot for a set of binary neutron star injections in gaussian noise. For unbiased recovery, the lines should lie perfectly along the diagonal in the above plot. The traces corresponding to all parameter trend along this diagonal, and are consistent with the error regions predicted due to finite number of injections (grey shaded region).}
\end{figure}

As a full parameter estimation of BNS signals with the exact method takes $\mathcal{O}(10^3)$ core hours, instead of comparing our posteriors with the exact method, we only remark that the injected parameters were well-recovered with respect to their true values. To test this further, we injected 143 BNS signals generated with the approximant \texttt{IMRPhenomPv2} into simulated Gaussian noise and created a percentile-percentile (p-p) plot ~\cite{Cook:2006} of the recoveries. The simulated BNSs were drawn uniformly in chirp mass $\mathcal{M}$ between $1.42 M_\odot$ and $1.62 M_\odot$, uniformly in mass ratio $q$ between $0.125$ and $1$, uniformly in individual aligned spin magnitudes $\chi_1$ and $\chi_2$ between $-0.05$ and $0.05$, and uniformly in sky location and orientation of the binary. Luminosity distances for the binaries were drawn assuming a prior $\propto D_L^2$ between 10 Mpc and 300 Mpc. The same priors are used for estimating parameters from these mock signals. For unbiased inference, the p-p plot (shown in Fig.~\ref{fig: BNS pp-test}) for each of the parameters should lie along the diagonal and should be consistent with error bars coming from finite number of samples. We see that this is the case for all the parameters that we consider, further illustrating that our implementation of relative binning produces unbiased posteriors.

\subsection{Non-quadrupolar Waveforms}
\label{sec:hm}
As GWs are a tensor field, they are most naturally decomposed in terms of $Y^{-2}_{lm}$, the spherical harmonic basis functions of spin-weight -2. For GWs from quasi-circular, non-precessing binaries with comparable masses, waveform templates that consist of only the dominant quadrupole modes ($\ell = 2$, $m = \pm 2$) are sufficient. However, sub-dominant modes play an important role in the detection and parameter estimation of BBHs with high mass ratios or precessing spins. Additionally, the effect of higher harmonics becomes more pronounced as the inclination angle changes from 0 to $\pi/2$ (see e.g. Refs.~\cite{Sathyaprakash2009, Varma2014, Mateu-Lucena:2021siq}). Relative binning, as described in Section \ref{sec:technical overview}, is designed for likelihood evaluations involving only the dominant mode. Using a different set of waveform ratio approximations for each component mode and modified binning schemes, the accuracy of likelihood can be increased for non-quadrupolar waveforms~\cite{Leslie:2021ssu, Narola:2023men}.

\begin{figure}[!htbp]
   \centering
\includegraphics[width=\linewidth]{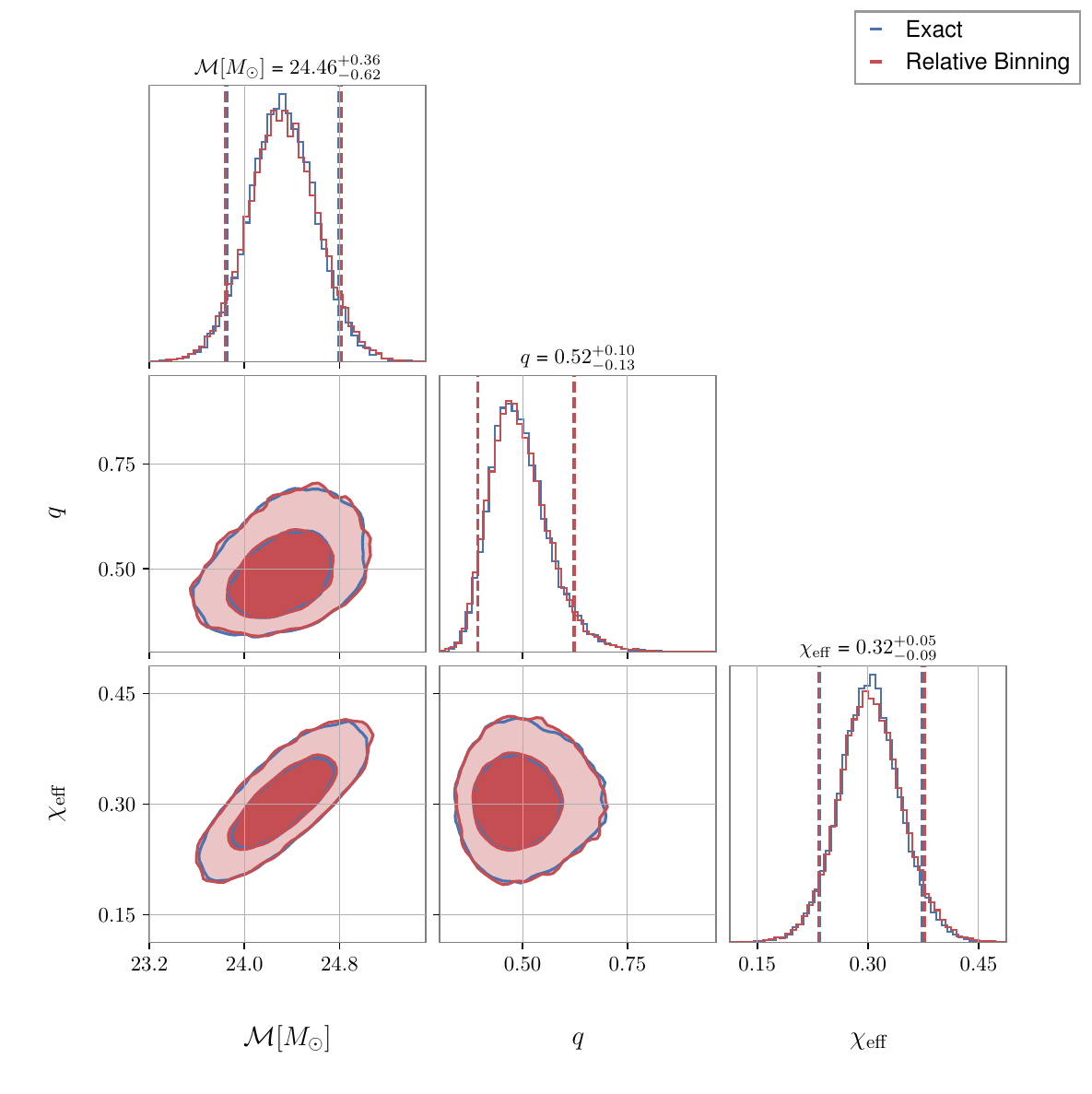} 
      \caption{Corner plot comparing the intrinsic parameters estimated using relative binning (red) and exact (blue) methods for a BBH injection with $m_{total} =120 M_\odot$ and $q=0.5$.
      The \texttt{IMRPhenomXPHM} approximant, which includes the effects of several non-quadrupole modes, was used for likelihood evaluations.}
      \label{fig: bbh_hm_intrinsic}
\end{figure}
\begin{figure*}[!htbp]
   \centering
   \includegraphics[width = 16cm]{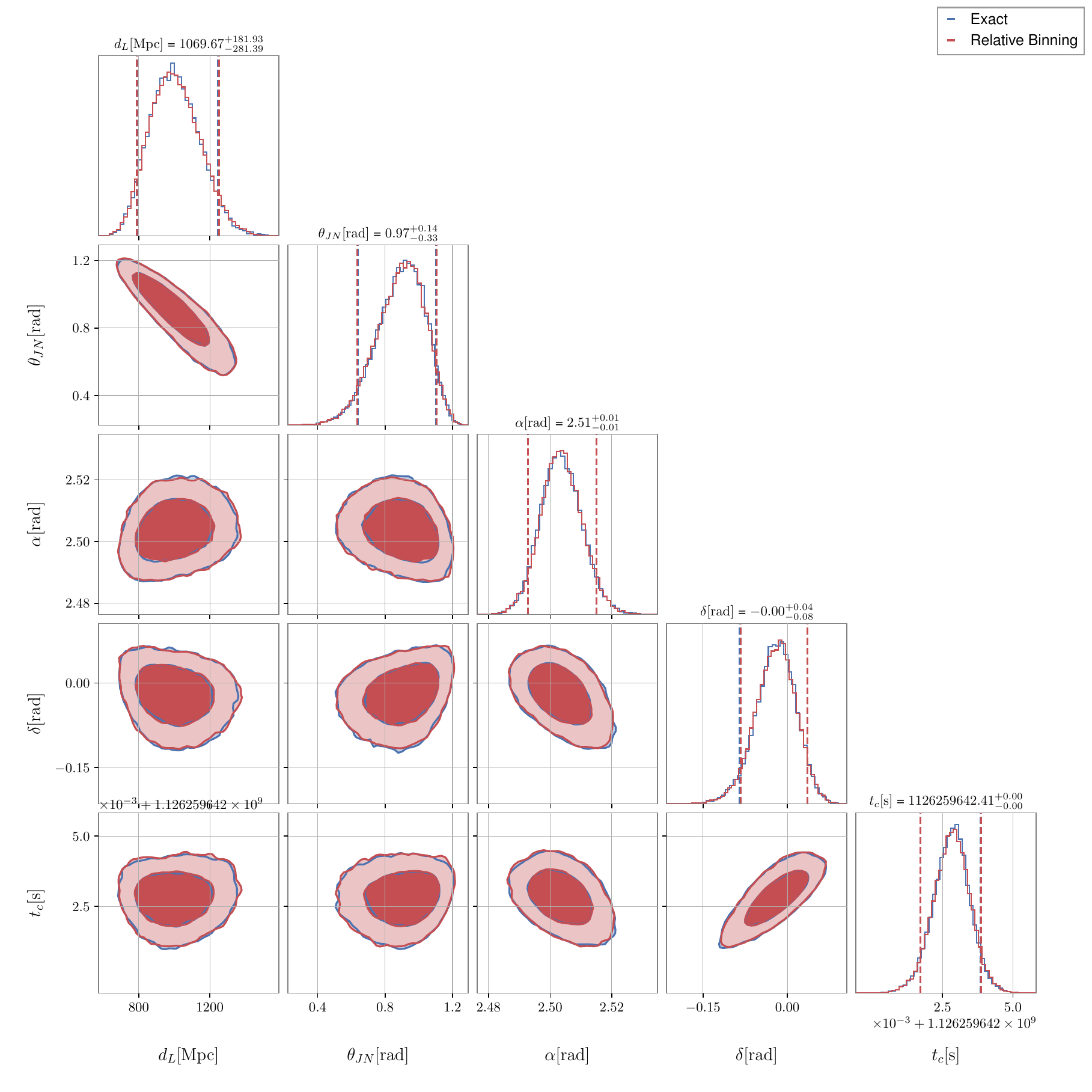} 
      \caption{Corner plot comparing the extrinsic parameters estimated using relative binning (red) and exact (blue) methods for the BBH injection seen in Figure \ref{fig: bbh_hm_intrinsic}.
      The \texttt{IMRPhenomXPHM} approximant, which includes the effects of several non-quadrupole modes, was used for likelihood evaluations.}
      \label{fig: bbh_hm_extrinsic}
\end{figure*}

\begin{figure}[!htbp]
   \centering
   \includegraphics[width=\linewidth]{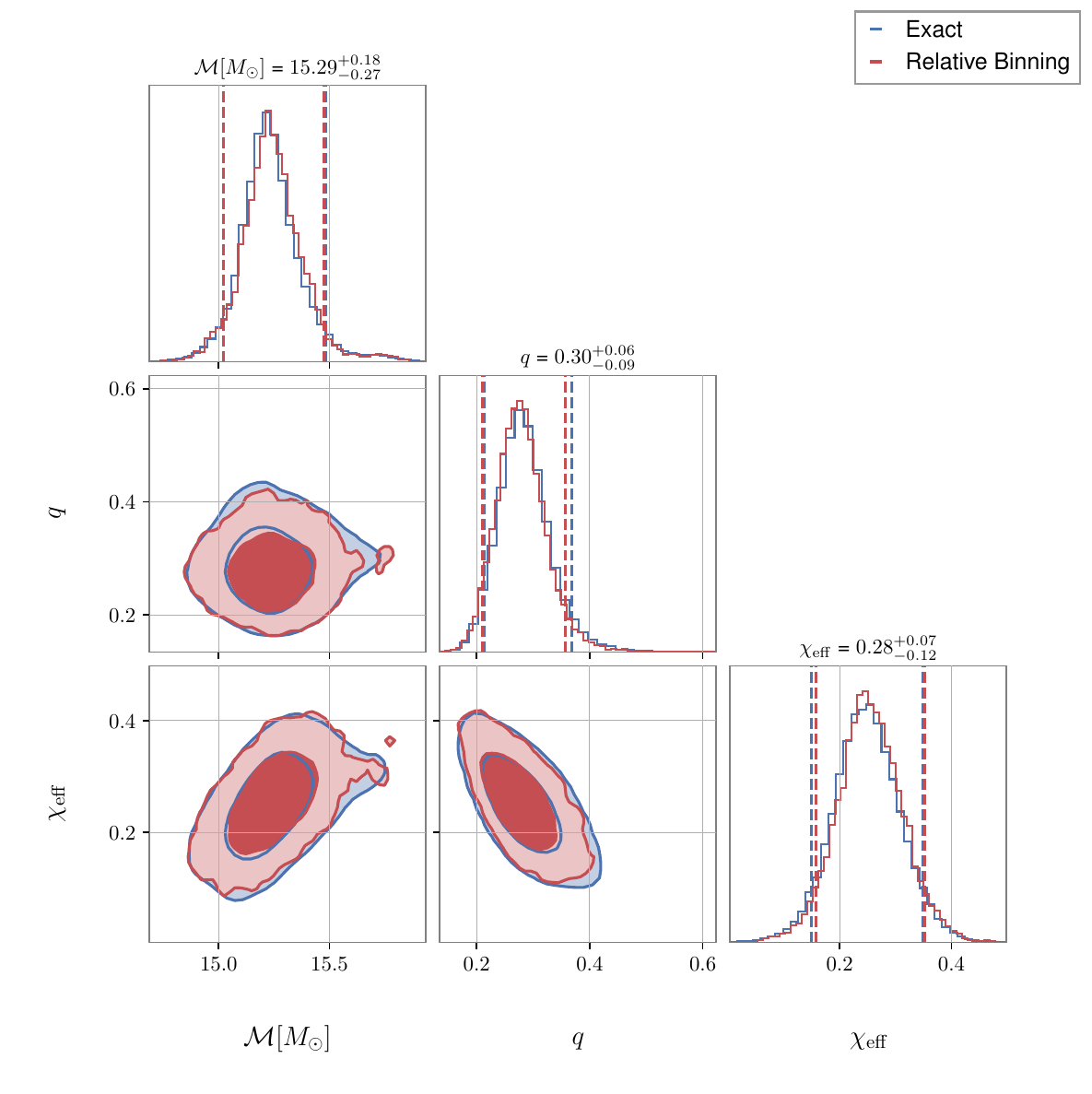} 
      \caption{Corner plot showing the posterior distributions on the intrinsic parameters of GW190412 calculated using the exact likelihood and the relative binning likelihood.}
 \label{fig:GW190412}
\end{figure}

 However, we tested our current implementation on BBHs with $60 M_\odot \leq m_{total} \leq 120 M_\odot$ and $0.25\leq q \leq1$ using the \texttt{IMRPhenomXPHM} approximant that includes the effects of non-quadrupole modes $(l, |m|) = (2, 2), (2, 1), (3, 3), (3, 2), (4, 4)$ and the mode mixing effects in $(l, |m|) = (3, 2)$ \cite{Pratten2020}. We find that the accuracy of posteriors is already quite good despite our code using the simplest prescription of relative binning with 123 frequency bins ($\epsilon = 0.25$). A comparison between the posteriors obtained with relative binning and the exact method for a BBH injection with $m_{total} =120 M_\odot$ and $q=0.5$ is shown in Figure \ref{fig: bbh_hm_intrinsic} and \ref{fig: bbh_hm_extrinsic}. The injection parameters are summarized in Table \ref{tab:injection-parameters}. The posteriors show good agreement with each other, with the maximum JS divergence being 0.0005 (corresponding to $\mathcal{M}$). 
 
 To further test our implementation against signals with significant higher mode content, we perform a parameter estimation of the event GW190412~\cite{LIGOScientific:2020stg} with relative binning. GW190412 is an asymmetric binary with a mass ratio $q \sim 1/4$, and contains strong evidence for higher order multipoles of GW radiation. We redo the exact-likelihood \texttt{Bilby} parameter estimation performed with the \texttt{IMRPhenomXPHM} approximant as a part of GWTC-2.1~\cite{LIGOScientific:2021usb}, but with the relative binning likelihood. All choices regarding the prior, sampling frequency, duration etc. are assumed to be exactly the same as that used for generating those samples. To ensure this, we use the \textsc{C01:IMRPhenomXPHM} configurations from the \texttt{*\_mixed\_nocosmo.h5} file corresponding to GW190412 from the GWTC-2.1 Zenodo data release~\cite{ligo_scientific_collaboration_and_virgo_2022_6513631} for posterior samples. We also choose our fiducial parameter value as the maximum likelihood sample in the released set of samples. The resulting comparison between samples (for the intrinsic parameters) obtained with the exact likelihood computation and the relative binning likelihood computation is shown in Fig.~\ref{fig:GW190412}. Again, the posteriors are in good agreement with each other, with a maximum JS divergence of 0.006 (corresponding to the primary spin magnitude $a_1$). This further showcases the ability of our implementation to recover parameters of signals with higher multipoles even though the prescription is not strictly optimal for such signals.
 
 We leave the \texttt{Bilby} implementation of relative binning specific to higher order modes to future work.

\subsection{Relative Binning for XG Detectors}
\label{sec:xg}

The next-generation (XG) ground-based detectors in the future such as the Einstein Telescope and Cosmic Explorer will have largely improved sensitivity compared to the Advanced LIGO and Virgo detector network. These detectors will have a frequency cutoff as low as $f_{\rm low}$=2Hz \cite{Punturo2011,LIGOScientific:2016wof}. We will observe currently detectable GW sources with a much higher SNR of the order $\mathcal{O}(10^3)$, thereby, allowing us to study these sources with greater precision \cite{Borhanian:2022czq, Gupta:2023lga, Branchesi:2023mws}. However, this would increase the sampling convergence time and make parameter estimation more expensive. Moreover, the inspiral signals can stay in the sensitivity band for a duration ranging between a few hours to days depending on the masses of the binary system \cite{Punturo2011}. To keep up with such a large amount of data generated, we need faster parameter estimation methods, and relative binning is a promising candidate.

In order to investigate the utility of our code XG detections, we injected a BBH and a BNS\footnote{The configuration used for the BNS signal (equal masses, source frame component mass of $2.4\,M_\odot$) is on the higher side when compared to predictions from standard equations of state of dense matter. While this configuration is probably unrealistic, our main conclusions about the utility of relative binning for XG detectors do not change.} signal with configurations corresponding to Einstein Telescope (ET) (assumed to be at the location of the Virgo detector in Italy) and Cosmic Explorer (CE) (two equally sensitive detectors at the location of LIGO-Hanford and LIGO-Livingston). For simplicity, we neglect the effect of a time-varying antenna pattern for the BNS injection. The power spectral densities used in this analysis are available\footnote{\url{https://git.ligo.org/lscsoft/bilby/tree/master/bilby/gw/detector/noise_curves}} in the \texttt{detector} module of the \texttt{gw} package in \texttt{Bilby}. We parallelize our calculations over 16 CPU cores. In both cases, the waveform approximant was \texttt{IMRPhenomPv2}, and the sampling frequency used was 4096 Hz. For the BBH signal, characterized by a signal duration of 32 seconds and injection parameters as detailed in Table \ref{tab:injection-parameters}, the SNR reaches 233, with the associated wall-clock time being 2 hours. On the other hand, for a BNS signal with a duration of 1024 seconds, a frequency range covering 10 to 2048 Hz, and an SNR of 313, the associated wall-clock sampling time was 3.5 hours. Table \ref{tab: 3g-signals} summarizes this information. We find that 123 frequency bins are sufficient even in this case.

\begin{table}[htb] 
\centering
\caption{Summary of XG ground-based detector injections}
\label{tab: 3g-signals}
\begin{tabular}{|c|c|c|c|} 
\hline
    &\begin{tabular}[c]{@{}c@{}}Signal Duration\\(seconds)\end{tabular} & \begin{tabular}[c]{@{}c@{}}Frequency Range\\(Hz)\end{tabular} & \begin{tabular}[c]{@{}c@{}}Sampling Time \\ (hours; 16 CPUs)\end{tabular}  \\ 
\hline
BBH & 32 & 6-2048 & 2 \\
\hline
BNS & 1024 & 10-2048 & 3.5\\
\hline
\end{tabular}
\end{table}

\begin{figure}[!htbp]
   \centering
   \includegraphics[width=\linewidth]{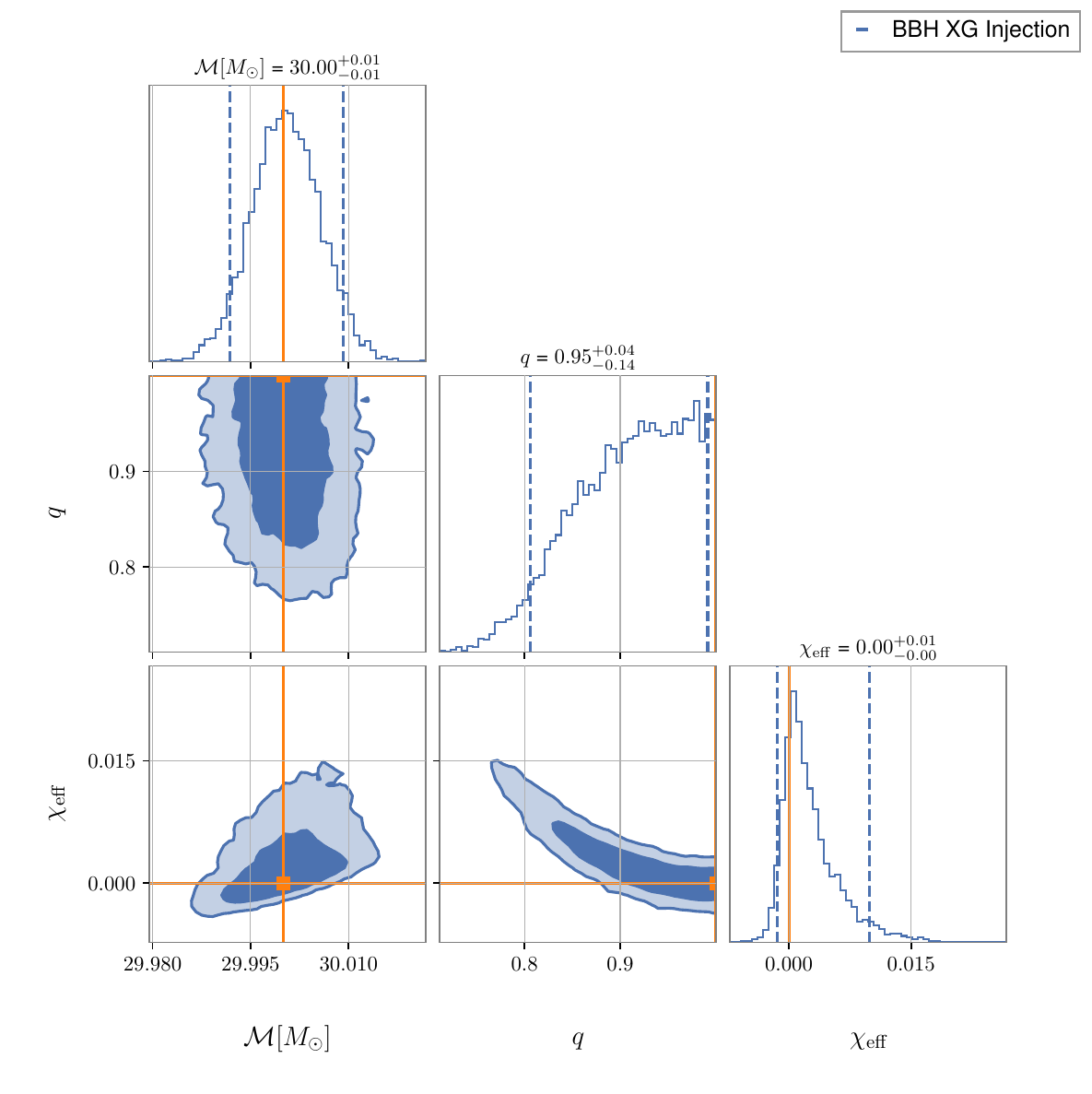} 
      \caption{Corner plot showing the posterior distributions for intrinsic parameters of a BBH injection in XG ground-based network estimated using relative binning.
      Solid orange lines indicate the injected parameters.
      The Marginalized 1-dimensional histograms for each parameter are shown along the diagonal with dashed vertical lines and labels for 5\% and 95\% quantiles of the parameters.
      Off-diagonal plots show 2-dimensional joint-posterior distributions with contours corresponding to the 68\% and 95\% credible regions. The waveform approximant used is \texttt{IMRPhenomPv2}. See table \ref{tab: 3g-signals} for more details.}
 \label{fig: bbh-3g-in}
\end{figure}

\begin{figure*}[!htbp]
   \centering
   \includegraphics[width=\linewidth]{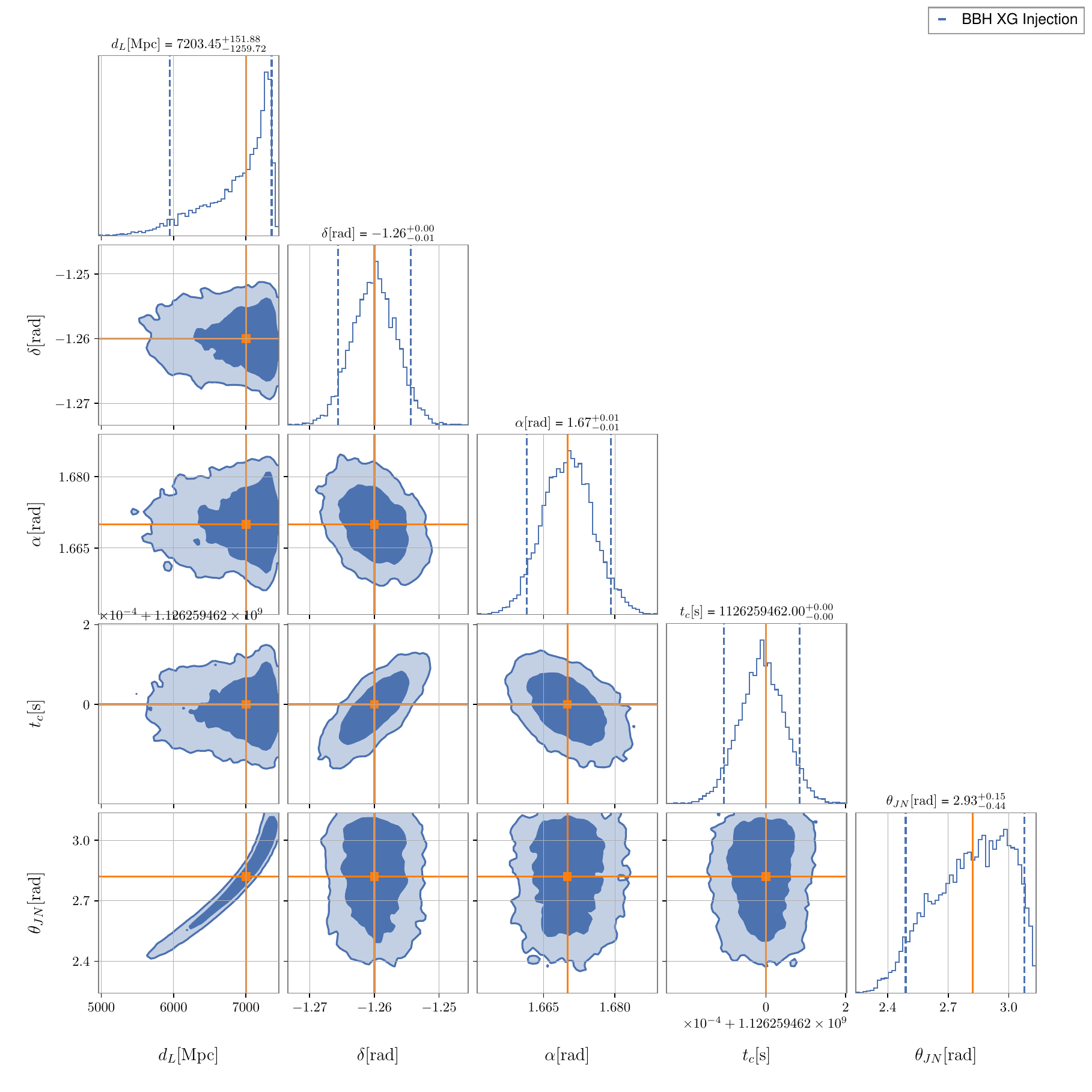} 
      \caption{Corner plot for extrinsic parameters of the BBH injection from Figure \ref{fig: bbh-3g-in}.
      Solid orange lines indicate the injected parameters.
      The Marginalized 1-dimensional histograms for each parameter are shown along the diagonal with dashed vertical lines and labels for 5\% and 95\% quantiles of the parameters.
      Off-diagonal plots show 2-dimensional joint-posterior distributions with contours corresponding to the 68\% and 95\% credible regions. The waveform approximant used is \texttt{IMRPhenomPv2}. }
      \label{fig: bbh-3g-ex}
\end{figure*}

\begin{figure}[htb!]
   \centering
\includegraphics[width=\linewidth]{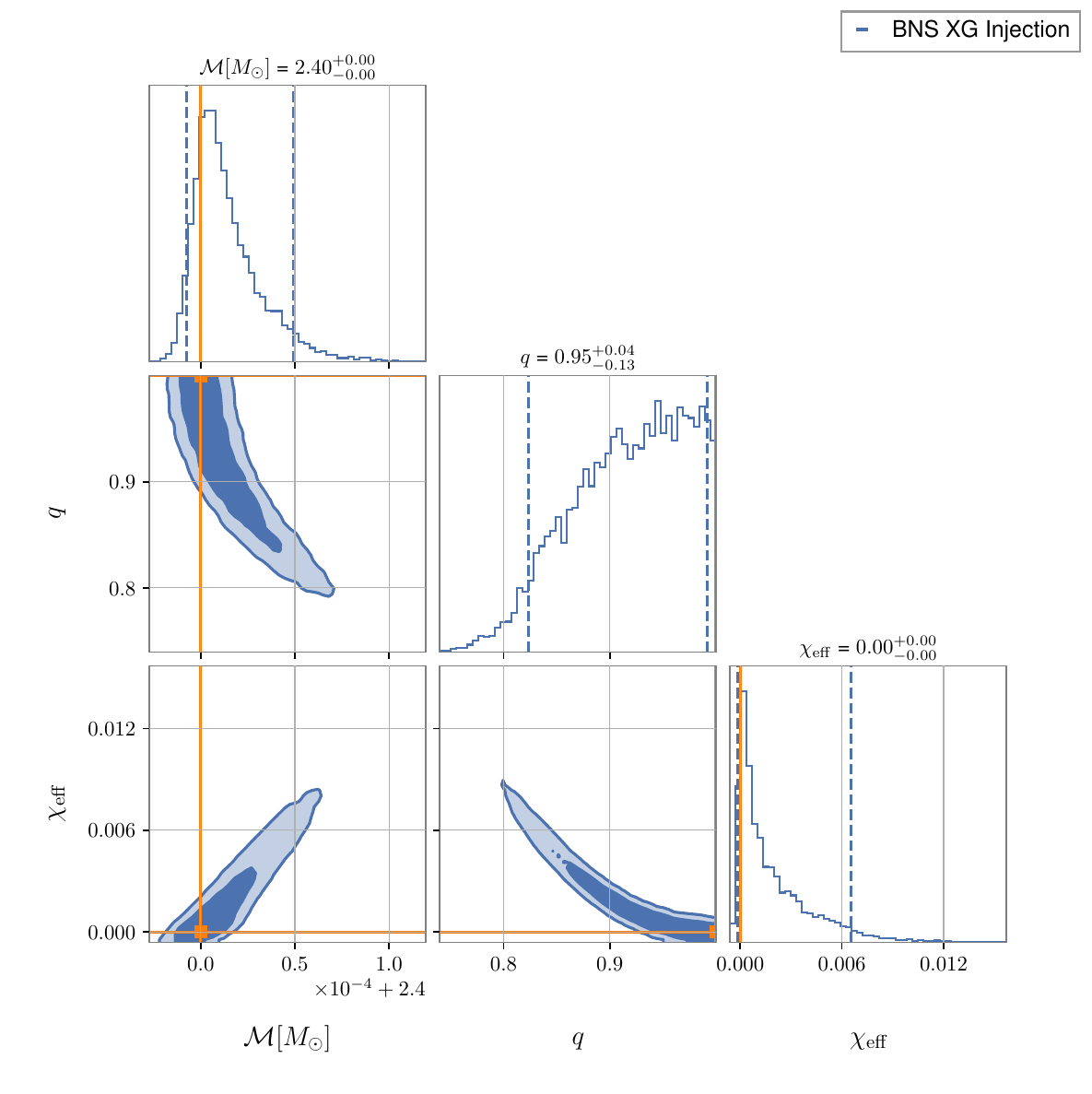} 
      \caption{Corner plot showing the posterior distributions for intrinsic parameters of a BNS injection in XG ground-based network estimated using relative binning.
      Solid orange lines indicate the injected parameters.
      The Marginalized 1-dimensional histograms for each parameter are shown along the diagonal with dashed vertical lines and labels for 5\% and 95\% quantiles of the parameters.
      Off-diagonal plots show 2-dimensional joint-posterior distributions with contours corresponding to the 68\% and 95\% credible regions. The waveform approximant used is \texttt{IMRPhenomPv2}. See table \ref{tab: 3g-signals} for more details.}
      \label{fig: bns-3g-in}
\end{figure}

\begin{figure*}[!htbp]
   \centering
   \includegraphics[width = \linewidth]{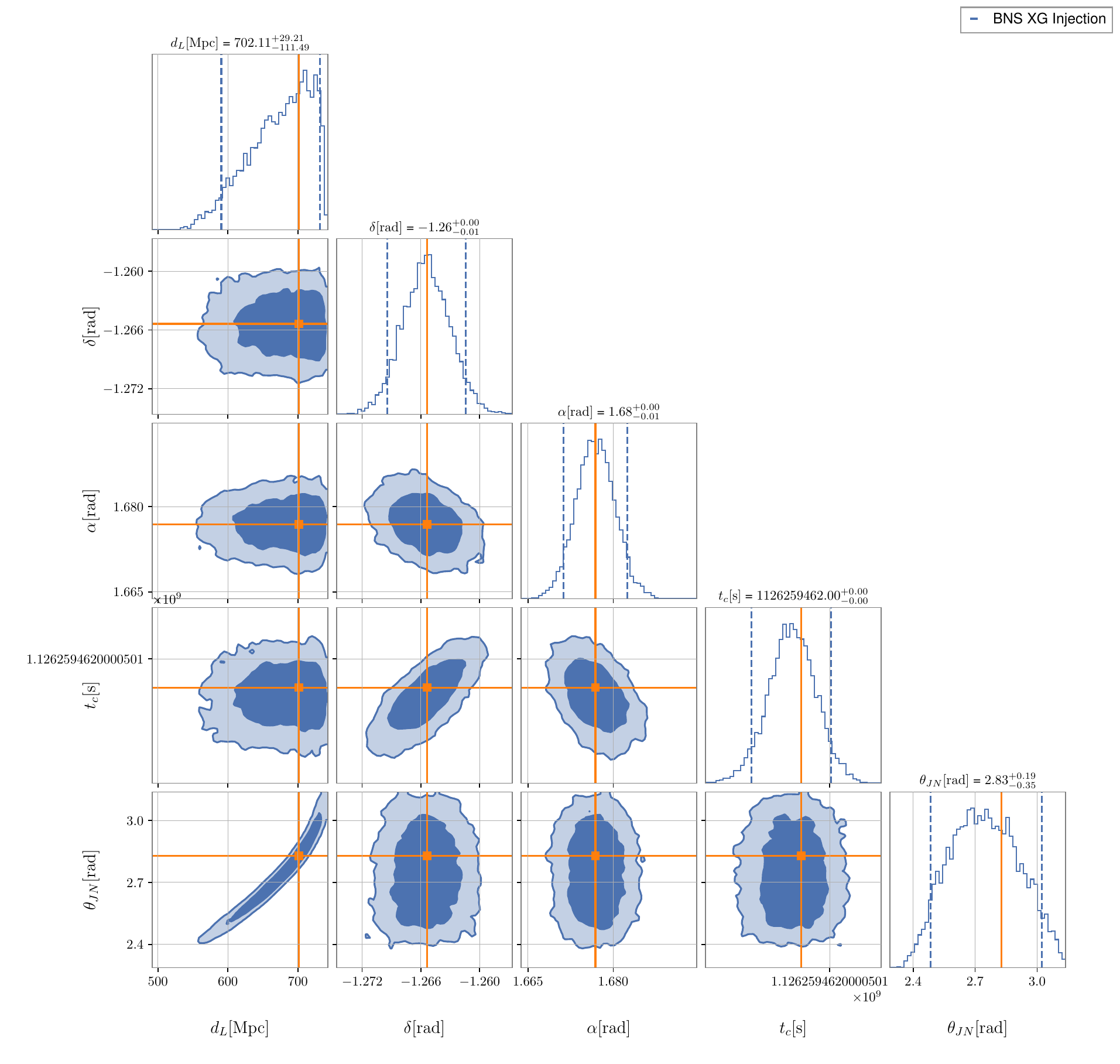} 
      \caption{Corner plot for extrinsic parameters of the BNS injection from Figure \ref{fig: bns-3g-in}.
      Solid orange lines indicate the injected parameters.
      The Marginalized 1-dimensional histograms for each parameter are shown along the diagonal with dashed vertical lines and labels for 5\% and 95\% quantiles of the parameters.
      Off-diagonal plots show 2-dimensional joint-posterior distributions with contours corresponding to the 68\% and 95\% credible regions. The waveform approximant used is \texttt{IMRPhenomPv2}.}
      \label{fig: bns-3g-ex}
\end{figure*}

The corner plots of these injections are shown in Figures \ref{fig: bbh-3g-in}, \ref{fig: bbh-3g-ex}, \ref{fig: bns-3g-in} and \ref{fig: bns-3g-ex}. We see that the posteriors of all the parameters are well-recovered for both signals. The sampling times reported here are comparable to the GW detections with Advanced LIGO/Virgo network despite the huge SNRs. Therefore, relative binning will be tremendously useful for parameter estimation of XG detections.

\subsection{Speed-Up Factors}

\begin{table*}[htb!] 
\centering
\caption{Speed-up factors for GW signals of different durations.
We used \texttt{IMRPhenomPv2} and \texttt{IMRPhenomPv2$\_$NRTidal} approximants for the BBH and BNS signals respectively.} \label{tab: speed-up} 
\begin{tabular}{|c|c|c|c|c|} 
\hline
\multirow{2}{*}{\textbf{GW Signal Type}} & \multirow{2}{*}{\textbf{Duration} (s)} & \multicolumn{2}{c|}{$\mathcal{L}$ \textbf{calc time} (ms) } & \multirow{2}{*}{\begin{tabular}[c]{@{}c@{}}\textbf{Speed-up }\\\textbf{Factor}\end{tabular}} \\ 
\cline{3-4}
 &  & \textbf{Exact} & \textbf{Relbin} &  \\ 
\hline
BBH (GW150914-like) & 4 & 3.967 & 1.245 & 3 \\ 
\hline
XG BBH & 32 & 55.538 & 1.663 & 33 \\ 
\hline
BNS (GW170817-like) & 128 & 275.89 & 1.713 & 161 \\ 
\hline
XG BBH & 1024 & 2069.99 & 1.643 & 1260 \\ 
\hline
XG BNS & 4096 & 9225.77 & 1.893 & 4874 \\
\hline
\end{tabular}
\end{table*}

The speed-up due to relative binning is two-fold.
Let us say $N$ is the number of points in a full frequency grid used for parameter estimation with the exact method. For relative binning, the
\begin{enumerate}[itemsep=0ex,topsep=0ex]
\item number of points at which a sampled waveform is generated goes as $\mathcal{O}${(number of bins)} instead of $\mathcal{O}${($N$)}.
\item number of overall complex-multiplications involved in likelihood computation goes as $\mathcal{O}${(number of bins)} instead of $\mathcal{O}${($N$)}.
\end{enumerate}

The relative speed-ups are more pronounced when analyzing long chunks of strain data with high sampling rates. For instance, an exact likelihood computation for a binary neutron star merger event like GW170817 with a duration of 256s and a sampling rate of 4096 Hz would require a frequency grid of the order $\sim10^5$. Whereas, with relative binning, we would only need bins $\sim100$. The speed-up factors for different signal lengths are tabulated in Table \ref{tab: speed-up}. Notice that the time per likelihood evaluation in relative binning is 1-2 milliseconds irrespective of the signal type or duration. This is the case because the number of frequency bins needed to obtain posteriors with the required accuracy remains more or less the same. In all our analyses, we have used only 123 bins. Hence, with an increase in signal duration the speed-up factor scales proportionally.

\section{Summary}\label{sec:summary}
Parameter estimation of GW signals is a computationally expensive process whose time complexity increases with growing signal durations. Relative binning is a promising candidate for speedy parameter estimation for future GW detections. In this work, we describe the implementation of relative binning in \texttt{Bilby}. We have shown that our implementation is able to successfully recover parameters of GW170817 in 14 hours on a single CPU core. The sampling time will reduce when the sampler is parallelized over multiple cores, as in done for the analyses in this work with XG detectors. We also showed that our implementation is able to recover parameters of injected signals, even for those with significant higher mode content. Our implementation passes p-p tests and  enables large-scale injection campaigns with low mass GW signals without very high associated computational costs. Specifically, relative binning is a promising candidate for analyzing CBCs in the XG era owing to its low computational cost.
Our \texttt{Bilby} implementation allows further testing and investigation of the technique and its use in the current observing run (O4) of the LVK collaboration. Additionally, \texttt{Bilby}'s user-friendly interface makes relative binning accessible to everyone.

Although we have performed validation checks by testing the accuracy of our code against the exact method for ample GW signals, more work needs to be done to assess its performance on a larger population. Such studies investigating the impact of the relative binning technique in various corners of the parameter space have been absent from the literature. We have shown that the advantage of using relative binning increases with duration of detected events. This makes it a powerful tool for parameter estimation of binary neutron star mergers and neutron star -- black hole mergers in O4. The promise of relative binning for multi-messenger parameter estimations and follow-ups has already been demonstrated~\cite{Raaijmakers2021,Finstad2020}. As we have shown, relative binning will also prove to be immensely useful for detections with XG ground-based interferometers.

\begin{acknowledgments}
We thank Ish Gupta and Ajith Parameswaran for a reading of our draft. AV thanks Ish Gupta and Divya Singh for many conversations about relative binning. We are also grateful to the entire \texttt{Bilby} O4 review team for their efforts in getting this pipeline reviewed for LVK analyses, especially Gregory Ashton, Charlie Hoy, and Simon Stevenson.

This work was supported by Department of Atomic Energy, Government of India, under Project No. RTI4001. AV acknowledges support from the Natural Sciences and Engineering Research Council of
Canada (NSERC) (funding reference number 568580).
CT is supported by the Eric and Wendy Schmidt AI in Science Postdoctoral Fellowship, a Schmidt Futures program. AZ was supported by NSF Grants PHY-2207594 and PHY-2308833.

This research has made use of data or software obtained from the Gravitational Wave Open Science Center (gwosc.org), a service of LIGO Laboratory, the LIGO Scientific Collaboration, the Virgo Collaboration, and KAGRA \cite{LIGOScientific:2019lzm}. LIGO Laboratory and Advanced LIGO are funded by the United States National Science Foundation (NSF) as well as the Science and Technology Facilities Council (STFC) of the United Kingdom, the Max-Planck-Society (MPS), and the State of Niedersachsen/Germany for support of the construction of Advanced LIGO and construction and operation of the GEO600 detector. Additional support for Advanced LIGO was provided by the Australian Research Council. Virgo is funded, through the European Gravitational Observatory (EGO), by the French Centre National de Recherche Scientifique (CNRS), the Italian Istituto Nazionale di Fisica Nucleare (INFN), and the Dutch Nikhef, with contributions by institutions from Belgium, Germany, Greece, Hungary, Ireland, Japan, Monaco, Poland, Portugal, Spain. KAGRA is supported by the Ministry of Education, Culture, Sports, Science and Technology (MEXT), Japan Society for the Promotion of Science (JSPS) in Japan; National Research Foundation (NRF) and the Ministry of Science and ICT (MSIT) in Korea; Academia Sinica (AS) and National Science and Technology Council (NSTC) in Taiwan.

This work has made use of the \texttt{numpy}~\cite{2020Natur.585..357H}, \texttt{scipy}~\cite{2020NatMe..17..261V}, \texttt{matplotlib}~\cite{2007CSE.....9...90H}, \texttt{jupyter}~\cite{2016ppap.book...87K}, \texttt{pandas}~\cite{mckinney-proc-scipy-2010}, \texttt{bilby}~\cite{Ashton:2018jfp}, \texttt{bilby\_pipe}~\cite{Romero-Shaw:2020owr}, \texttt{corner}~\cite{corner}, \texttt{dynesty}~\cite{Speagle:2019ivv}, \texttt{PESummary}~\cite{Hoy:2020vys} software packages.

\end{acknowledgments}

\appendix
\section{Time Marginalization in Relative Binning} \label{sec:time-marg}
In this appendix, we demonstrate why time marginalization is not feasible for relative binning. The formula for time-marginalised likelihood \cite{Thrane2019} is
\begin{equation}\label{eq: time-marg-lnlike}
\log \mathcal{L}_\text{marg}^t = \log{\cal Z}_N - \frac{1}{2} \rho_\text{opt}^2(\bm{\tilde{\theta}}) + \log \sum_{n=0}^{N-1} e^{\kappa^{2}(\bm{\tilde{\theta}}, n)} \pi_n 
\end{equation}
where
\begin{equation}\label{eq: kappa-fft}
\kappa^{2}(n) = 4 \Delta f  \Re \Big\{ {\tt fft}_{n}\left(\frac{d \ \mu^{*}(t_0)}{S_{n}(f)} \right) \Big\}
\end{equation}
With relative binning, the computation of the $\kappa^2(n)$ array is not that straightforward. Let us consider two possible ways to do this.
\begin{itemize}
\item Alternative I:
In Table \ref{tab:fvsbin}, we saw that the waveform ratio $r(f)$ for any given waveform $\mu(\bm{\tilde{\theta}}, f)$ is only evaluated at the bin edges.
But let's say we want the values of $r(f)$ over the frequency grid.
One way to do this would be to interpolate the existing values of $r(f)$ using the linear-fit coefficients (which were obtained from $r(f)$ at the bin edges) as seen in Eq.
\ref{eq: r linear}.
\begin{equation}\label{eq:}
r_{interp}(f)=r_{0}(\mu, \mathrm{b})+r_{1}(\mu, \mathrm{b})\left(f-f_{\mathrm{m}}(\mathrm{b})\right)
\end{equation}
Now the interpolated waveform ratio that spans the entire frequency grid can be used to compute the $\kappa^2(n)$ array as follows:
\begin{equation}\label{eq: }
\begin{aligned}
\kappa^{2}(n) &= 4 \Delta f  \Re \Big\{ {\tt fft}_{n}\left(\frac{d \ \mu^{*}(t_0)}{S_{n}(f)} \right) \Big\} \\
             &= 4 \Delta f  \Re \Big\{ {\tt fft}_{n}\left(\frac{d \ \mu_0^{*}(t_0) r^{*}_{\rm interp}(t_0)}{S_{n}(f)} \right) \Big\}
\end{aligned}
\end{equation}
where we have used the fact that $\mu(f) = r(f) \mu_0(f)$. This is very similar to the time marginalization implemented with the ROQ likelihood in \texttt{Bilby}. 

\item Alternative II

The other alternative is to modify the summary data corresponding to the matched filter.
We start with eq.
\ref{eq: kappa-fft} and replicate the summary data derivations in section \ref{sec: sd}.

\begin{widetext}
\begin{align}
\kappa^{2}(n) &= 4\Delta f \Re \sum_{j}^{N} \frac{d_{j} \mu_j^{*}( t_0)}{S_{n}(f_j)} e^{-2 \pi i j \frac{n}{N-1}} \nonumber \\
            &= \Re \sum_{b} \biggl( 4 \sum_{f_i \in b} \dfrac{d(f_i) r^*(f_i) \, \mu_{0}^*(f_i, \bm{\tilde{\theta}}) } {S_n(f_i) \ T} e^{-2 \pi i j \frac{n}{N}} \biggr)   \nonumber\\
            &\approx \Re \sum_{b} \Biggl(  r_{0}^{*}(\mu, \mathrm{b}) \biggl(4 \sum_{f_i \in b} \dfrac{d(f_i) \ \mu_{0}^*(f_i, \bm{\tilde{\theta}}) } {S_n(f_i)\ T} e^{-2 \pi i j \frac{n}{N}}\biggr)  \nonumber \\
            &\qquad \qquad+ r_{1}^{*}(\mu, \mathrm{b}) \biggl(4 \sum_{f_i \in b} \dfrac{d(f_i) \ \mu_{0}^*(f_i , \bm{\tilde{\theta}}) } {S_n(f_i)\ T} (f-f_{\mathrm{m}}(\mathrm{b})) e^{-2 \pi i j \frac{n}{N}} \biggr) \Biggr)  \nonumber\\
            &\approx \Re \sum_{\mathrm{b}}\left(A^{t}_{0}(\mathrm{b}) r_{0}^{*}(\mu, \mathrm{b})+A^{t}_{1}(\mathrm{b}) r_{1}^{*}(\mu, \mathrm{b})\right) \\
\mathrm{where} \qquad A^{t}_{0}(n, \mathrm{b}) &=4 \sum_{f \in \mathrm{b}} \frac{d(f) \mu_{0}^{*}(f)}{S_{n}(f) \ T}e^{-2 \pi i j \frac{n}{N}} \\
A^{t}_{1}(n, \mathrm{b}) &=4 \sum_{f \in \mathrm{b}} \frac{d(f) \mu_{0}^{*}(f)}{S_{n}(f) \ T} e^{-2 \pi i j \frac{n}{N}} \left(f-f_{\mathrm{m}}(\mathrm{b})\right)
\end{align}
\end{widetext}

We end up with modified summary data $A^{t}_{0}(n, \mathrm{b})$ and $A^{t}_{1}(n, \mathrm{b})$. Note that these terms also depend on $n$, the discretised coalescence time. On the other hand, $B_{0}(\mathrm{b})$ and $B_{1}(\mathrm{b})$ remain unchanged as $\rho^2_\text{opt}(\bm{\tilde{\theta}})$ is unaffected by the coalescence time.
\end{itemize}
In Alternative I, though the waveform generation time is saved, the number of complex multiplications involved goes as $\mathcal{O}(N)$, where $N$ is the number of points in a full FFT grid.
Therefore, the marginalised likelihood computation takes more time than the normal likelihood computation with relative binning in this case. Whereas in Alternative II, though the two-fold time saving discussed in Section \ref{sec:technical overview} still holds, the pre-computation time is increased by  $\mathcal{O}(2N)$. This is because $A^{t}_{0}(n, \mathrm{b})$ and $A^{t}_{1}(n, \mathrm{b})$ will have to be calculated at $N$ points for each bin as $n = 0, 1, ..., N-1$. In this case, the increase in the pre-computation time is higher than the time saved by reducing the dimensionality of the parameter space by one with time marginalization. Therefore, with relative binning,  marginalizing over time only slows down the parameter estimation. Nevertheless, Alternative I is included as an option for time marginalization in the \texttt{Bilby} implementation.

\bibliography{references}
\end{document}